\documentclass[aps,showpacs,nofootinbib,superscriptaddress,twocolumn]{revtex4}

\usepackage{graphicx}
\usepackage{bm}
\usepackage{amsmath}
\usepackage{amssymb}
\usepackage{hyperref}

\newcommand{\xbj}{x}
\newcommand{\qslash}{\kern 0.2 em n\kern -0.50em /}
\newcommand{\nslash}{\kern 0.2 em n\kern -0.50em /}
\newcommand{\kslash}{\kern 0.2 em k\kern -0.45em /}
\newcommand{\lslash}{\kern 0.2 em l\kern -0.50em /}
\newcommand{\pslash}{\kern 0.2 em p\kern -0.50em /}
\newcommand{\Sslash}{\kern 0.2 em S\kern -0.50em /}
\newcommand{\Pslash}{\kern 0.2 em P\kern -0.50em /}
\newcommand{\Dslash}{\kern 0.2 em D\kern -0.65em /\kern 0.15em}

\newcommand{\bp}{\boldsymbol{p}_T}
\newcommand{\bP}{\boldsymbol{P}_T}
\newcommand{\bk}{\boldsymbol{k}_T}

\newcommand{\ssh}{\!\!\!/}
\newcommand{\Tr}{\operatorname*{Tr}\nolimits}

\newcommand{\ph}{\phi_h}

\begin{document}

\title{ Transverse single-spin asymmetries of pion production in semi-inclusive DIS at subleading twist}

\author{Wenjuan Mao}\affiliation{Department of Physics, Southeast University, Nanjing
211189, China}
\affiliation{School of Physics and State Key Laboratory of Nuclear Physics and Technology, Peking University, Beijing
100871, China}
\author{Zhun Lu}\email{zhunlu@seu.edu.cn}\affiliation{Department of Physics, Southeast University, Nanjing
211189, China}
\author{Bo-Qiang Ma}\email{mabq@pku.edu.cn}\affiliation{School of Physics and State Key Laboratory of Nuclear Physics and Technology, Peking University, Beijing
100871, China}

\begin{abstract}
We study the single-spin asymmetries with the $\sin\phi_S$
and $\sin(2\phi_h -\phi_S)$ angular dependences for charged and neutral pions produced in semi-inclusive deep inelastic scattering on the transversely polarized proton target.
The theoretical interpretations of the two asymmetries are presented in terms of the convolution of the twist-3 quark transverse momentum dependent distributions and twist-2 fragmentation functions.
Specifically, we investigate the role of the distributions $f_T$, $h_T$ and $h_T^\perp$ in the $\sin\phi_S$ asymmetry, as well as the role of the distributions $f_T^\perp$, $h_T$ and $h_T^\perp$ in the $\sin(2\phi_h -\phi_S)$ asymmetry.
We calculate these distributions in a spectator-diquark model and predict the corresponding asymmetries for the first time, considering the kinematics at HERMES, JLab and COMPASS.
The numerical estimates show that the asymmetries are sizable, and the dominant contribution to the $\sin\phi_S$ asymmetry comes from the T-odd distribution $f_T$, while $f_T^\perp$ gives the main contribution to the $\sin(2\phi_h -\phi_S)$ asymmetry.
The future measurements on these asymmetries can shed light on the information of twist-3 transverse momentum dependent distributions.
\end{abstract}


\maketitle

\section{Introduction}

Transverse spin phenomena of the nucleon in scattering processes have received a lot of attention in recent years since they provide new insights on the understanding of hadron structure (for reviews see \cite{bdr,D'Alesio:2007jt,Barone:2010ef}).
Particularly, an ideal tool to probe the transverse spin structure of the nucleon is the single-spin asymmetry (SSA) in semi-inclusive deep inelastic scattering (SIDIS) off the transversely polarized target,
as reflected in numerous studies from both theory and experiment.
The typical transverse SSAs appearing in SIDIS are the Sivers asymmetry~\cite{Boer:1997nt,Brodsky:2002cx} and the Collins asymmetry~\cite{Collins:1992kk}.
The Sivers asymmetry has a $\sin(\phi_h-\phi_S)$ modulation, where $\phi_h$ and $\phi_S$ are the azimuthal angles of the detected hadron and the nucleon spin with respect to the lepton scattering plane.
It involves the convolution of the Sivers function~\cite{Sivers:1989cc,Anselmino:1994tv} that is a T-odd transverse momentum dependent (TMD) distribution function, and the ordinary unpolarized fragmentation function (FF).
The Collins asymmetry has a $\sin(\phi_S+\phi_h)$ azimuthal angular dependence, and can be interpreted in terms of
the transversity distribution function~\cite{Ralston:1979ys}, combined with the Collins FF~\cite{Collins:1992kk}.
A remarkable feature of the above mentioned asymmetries is that they appears at leading twist, thereby those effects should not be unsuppressed at high energies.
Indeed significant asymmetries were measured by the HERMES Collaboration~\cite{Airapetian:2004tw,Airapetian:2009ae,
Airapetian:2010ds}, the COMPASS Collaboration~\cite{Alexakhin:2005iw,Ageev:2006da,Alekseev:2008aa,
Alekseev:2010rw,Adolph:2012sn,Adolph:2012sp}
and Jefferson Lab (JLab) Hall A Collaboration~\cite{Qian:2011py,Zhao:2014qvx}.
The corresponding data have been utilized to extract~\cite{Efremov:2004tp,Vogelsang:2005cs,
Anselmino:2012aa,Anselmino:2013vqa} the Sivers function and transversity distribution, within the TMD factorization~\cite{Ji:2004wu}. At leading twist there is another transverse SSA, related to the pretzelosity distribution~\cite{Avakian:2008dz,She:2009jq,Lorce:2011kn}, and it was measured by the Hall A Collaboration~\cite{Zhang:2013dow} through its characteristic $\sin{(3\phi_h-\phi_S)}$ moment very recently.

However, the leading-twist SSAs do not exhaust all possible azimuthal dependences in SIDIS off a transversely polarized target from an unpolarized lepton beam. As shown in Refs.~\cite{Diehl:2005pc,Bacchetta:2006tn}, theoretically there are two more angular modulations (assuming one photon exchange), the $\sin\phi_S$ and the $\sin(2\phi_h-\phi_S)$ moments, appearing in the the process $l N^\uparrow \rightarrow l^\prime + h+ X$.
Experimentally there is also an attempt~\cite{Parsamyan:2013ug} to measure those asymmetries.
The leading-twist dynamics cannot account for them.
According to the analysis in Ref.~\cite{Bacchetta:2006tn}, in the partonic picture, these asymmetries can be explained by the convolution of various twist-3 distribution/fragmantation functions with twist-2 fragmentation/distribution functions.
Although the transverse spin asymmetries from the dynamical subleading-twist effects can shed light on the transverse spin structure of the nucleon at twist 3, there are still less systematic studies and calculations on the $\sin{\phi_S}$ and $\sin{(2\phi_h-\phi_S)}$ asymmetries in literature, especially from the phenomenological point of view.
We notice that sizable spin asymmetries related to subleading-twist dynamics have already been measured in other SIDIS processes, such as the longitudinal-target SSA~\cite{Airapetian:2005jc}, as well as the longitudinal-beam SSA~\cite{clas04,hermes07,Aghasyan:2011ha,Gohn:2014zbz}.
Therefore it will be quite necessary to investigate also the roles of twist-3 TMD distributions and FFs in the transverse SSAs, and to study the feasibility of experimental measurements on them, which are the main purpose of this work.

Both twist-3 distributions and FFs could give rise to the transverse SSAs.
In this paper, we will focus particularly on the contributions from twist-3 distributions.
We note that in the common reference frame~\cite{Bacchetta:2004jz} used to analyze SIDIS, the interaction-dependent twist-3 FFs (denoted with a tilde) also appear in the convolution.
In practical calculation these FFs may be set to zero in the Wandzura-Wilczek approximation~\cite{Wandzura:1977qf}.
However, recent studies~\cite{Kang:2010zzb,Metz:2012ct,Kanazawa:2014dca} on the contributions of the chirally and time-reversal odd FFs to the SSA in proton-proton collisions within the collinear twist-3 factorization, show that the fragmentation contributions from three parton correlation could still be sizeable.
In the light of those studies, it is possible that the contributions in SIDIS to the
$\sin\phi_S$ and $\sin(2\phi_h-\phi_S)$ asymmetries from the chiral- and T-odd FF $\tilde{H}$ might also be non-negligible.
Although it is interesting to investigate the effect of $\tilde{H}$ to the SSAs in SIDIS, the connection between the TMD-type FF and the collinear-type FF at the twist-3 level is still not clear.
Therefore, in this work we will not consider the fragmentation contributions, but will remain them as a future study.
%This approximation, especially for the T-odd twist-3 FFs, may be evident from the calculation by a spectator %model~\cite{Gamberg:2008yt}, as well as a model-independent analysis~\cite{Metz2008prl} on the T-odd collinear %quark-gluon-quark
%correlators, showing that the gluonic (partonic) pole contributions to fragmentation functions vanish.
%Therefore we expect that the contribution from the interaction-dependent twist-3 FFs is negligible.
In this scenario, then four twist-3 TMD distributions are involved in the transverse SSAs: $f_T$, $f_T^\perp$, $h_T$ and $h_T^\perp$.
The first one contributes to the $\sin\phi_S$ asymmetry, while the second one contributes to the $\sin{(2\phi_h-\phi_S)}$ asymmetry; the last two distributions contribute to both asymmetries through the convolution with the Collins FF.

The remained content of the paper is organized as follows.
In Section II, we calculate the TMD distributions $f_T$, $f_T^\perp$, $h_T$ and $h_T^\perp$ for the $u$ and $d$ valence quarks, as it is necessary to know their magnitudes and signs to predict SSAs.
As a demonstration we will use the spectator-diquark model developed in Ref.~\cite{Bacchetta:2008af}, which is also applied in Ref.~\cite{Mao:2012dk,Mao:2013waa}.
In Section III, using the model results obtained in Section II, we present our prediction on the $\sin\phi_S$ and $\sin(2\phi_h -\phi_S)$ asymmetries for charged and neutral pions in SIDIS, considering experimental configurations accessible at HERMES, JLab and COMPASS.
Although the TMD factorization at twist-3 level has not been proved~\cite{Gamberg:2006ru,Bacchetta:2008xw}, here we would like to adopt a more phenomenological way, i.e., to use the tree level result in Ref.~\cite{Bacchetta:2006tn} to perform the estimate.
Finally, we give our conclusion in Section IV.
Sect.~\ref{functions}

\section{Calculation of twist-3 TMD distributions in spectator-diquark model}
\label{functions}
\begin{figure}
  \includegraphics[width=0.8\columnwidth]{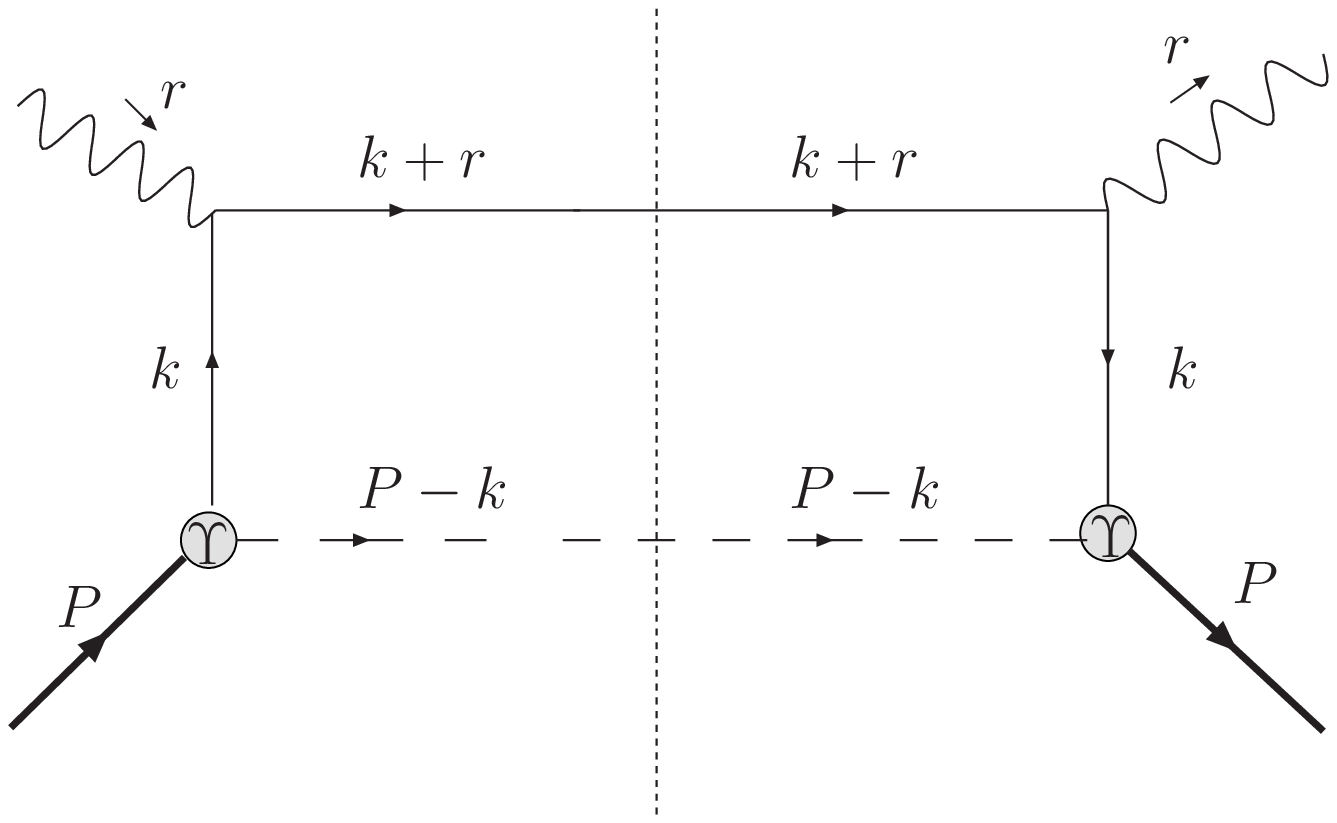}\\
  \includegraphics[width=0.8\columnwidth]{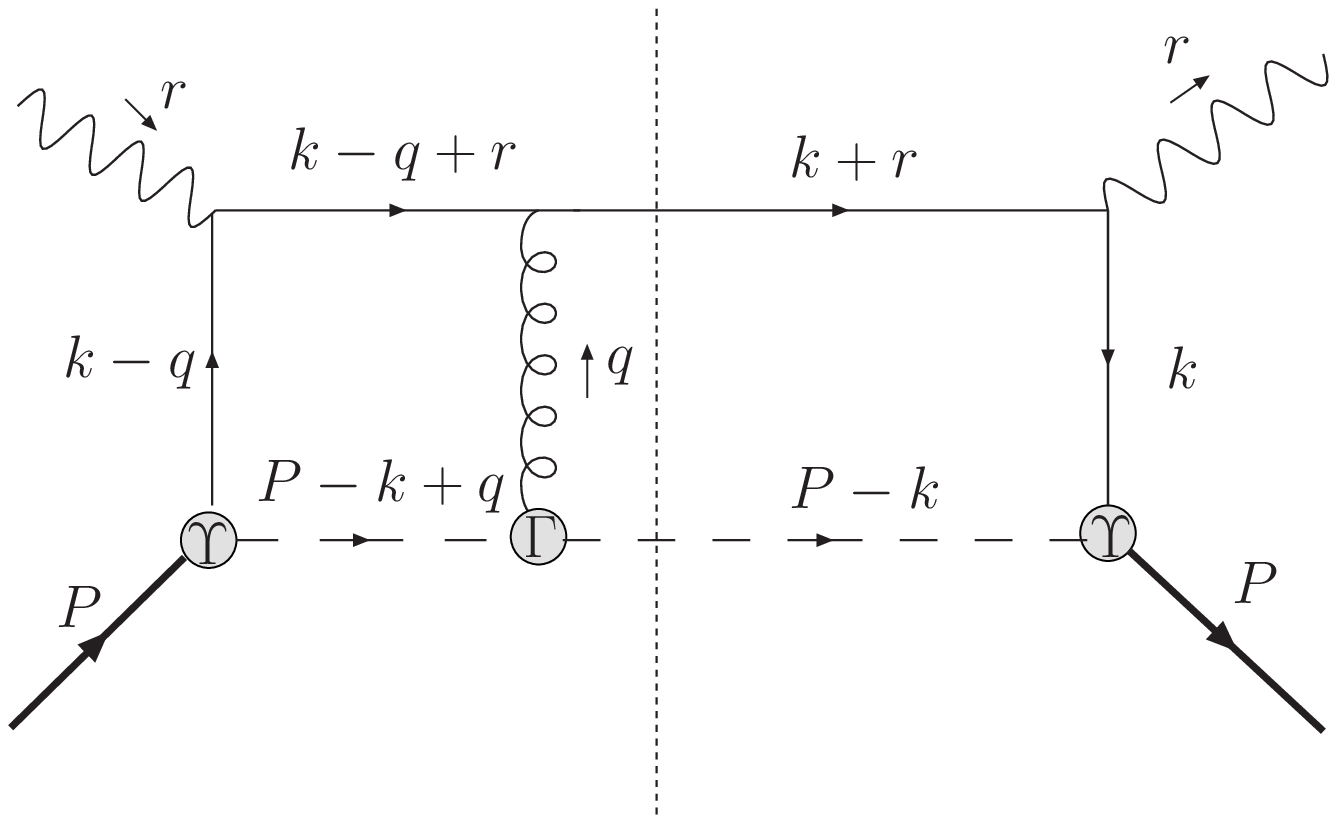}
 \caption{Cut diagrams for the spectator model calculation at tree level (upper) and one-loop level (lower). The dashed lines denote the spectator-diquarks that can be scalar diquarks or axial-vector diquarks.}
 \label{diagrams}
\end{figure}

In this section, we present the calculation on the four twist-3 TMD distributions in a spectator model, which was developed in Ref.~\cite{Bacchetta:2008af}.
In this model, the proton is supposed to be constituted by a quark and a diquark, and the diquark can be a scalar particle or an axial-vector one.
The relevant diagrams for the calculation are shown in Fig.~\ref{diagrams}, which are identical for the scalar and axial-vector cases.

The gauge-invariant quark-quark correlator can be expressed as
\begin{align}
\Phi(x,\bm k_T)&=\int {d\xi^- d^2\bm{\xi}_T\over (2\pi)^3}e^{ik\cdot\xi}
\langle PS|\bar{\psi}_j(0)\mathcal{L}[0^-,\infty^-]\nonumber\\
& \times \mathcal{L}[\bm 0_T,\bm \xi_T]\mathcal{L}[\infty^-,\xi^-]\psi_i(\xi)|PS\rangle\,.
\label{Phi}
\end{align}
For convenience here we adopt the light-cone coordinates $[a^-,a^+,\bm a_T]$ for an arbitrary four-vector $a$, with $a^{\pm}=(a^0\pm a^3)/\sqrt{2}=a\cdot n_{\mp}$, where the two light-like vectors are defined as $n_+=[0,1,\bm{0}_T]$ and $n_-=[1,0,\bm{0}_T]$.
The vector $\bm a_T=[a^1,a^2]$ denotes the two-component transverse vector that is perpendicular to the vectors $n_{\pm}$.
It is often to promote $\bm a_T$ to a four-vector
$a_T = [0,0,\bm a_T]$, and the scalar product of two transverse four-vectors satisfies
\begin{align}
a_T \cdot b_T = - \bm a_T \cdot \bm b_T.
\end{align}

At twist-3 level, the correlator (\ref{Phi}) for a transversely polarized nucleon can be decomposed into~\cite{Bacchetta:2006tn}:
\begin{align}
\Phi(x,\bm k_T,\bm S_T)\bigg{|}_{\textrm{twist-3}}
&= {M\over 2P^+}\left\{-\epsilon_T^{\rho\sigma}\gamma_\rho S_{T\sigma} f_T^\prime \right.\nonumber\\
+&\left.\frac{(k_T \cdot {S}_T)\epsilon_T^{\rho\sigma}\gamma_\rho k_{T\sigma}}{M^2} f_T^{\perp}\right.\nonumber\\
-&\left.\frac{k_T \cdot {S}_T}{M}\frac{[\nslash_+,\nslash_-]\gamma_5}{2} h_T \right.\nonumber\\
+&\left.\frac{[\Sslash_T,\kslash_T]\gamma_5}{2M} h_T^\perp+\cdots\right\},\label{phist}
\end{align}
here $\cdots$ denotes other twist-3 distributions that are not relevant in our calculation.
In the above decomposition we apply the notation from Ref.~\cite{Bacchetta:2006tn}.
We note that a different notation has been used in Ref.~\cite{Goeke:2005hb}.
The two T-even distributions $h_T$ and $h_T^\perp$, introduced in Ref.~\cite{Mulders:1995dh}, have been calculated in the spectator-diquark model~\cite{Jakob:1997wg} and the bag model~\cite{Avakian:2010br}.
The T-odd distributions $f_T$ is proposed in Ref.~\cite{Boer:1997nt}, while another T-odd distribution $f_T^\perp$ is a new function introduced in Ref.~\cite{Goeke:2005hb}.
They can be viewed as the analogy of the Sivers function at twist-3 level, and have been studied in Refs.~\cite{Gamberg:2006ru,Lu:2012gu} in scalar-diquark models.

We can obtain the twist-3 TMD distributions $f_T$, $f_T^\perp$, $h_T$ and $h_T^\perp$ from the correlator $\Phi(x,\bm k_T,\bm S_T)$ by the traces below:
\begin{align}
\frac{1}{2}\Tr[\Phi\,\gamma^{\alpha}]&=\frac{M}{P^+}\left[-\epsilon_T^{\alpha\rho} S_{T\rho} f_T^\prime\right.\nonumber\\
&+\left.\frac{(k_T \cdot  {S}_T)\epsilon_T^{\alpha\rho}k_{T\rho}}{M^2}f_T^{\perp}\right]\nonumber\\
&=\frac{M}{P^+}\left[-\epsilon_T^{\alpha\rho} S_{T\rho} f_T\right.\nonumber\\
-&\left.\frac{(k_T^\alpha k_T^\rho -{1\over 2} \,k_T^2 g_T^{\alpha\rho})}{M^2}\epsilon_{T\rho\sigma}S_T^{\sigma}f_T^{\perp}\right],
\label{fTs}\\
 \frac{1}{2}\Tr[\Phi\, i\sigma^{+-}\gamma_5]&=-\frac{M}{P^+}\left[\frac{k_T\cdot {S}_T}{M}h_T\right],
 \label{hT}\\
\frac{1}{2}\Tr[\Phi \,i\sigma^{\alpha\beta}\gamma_5]&=\frac{M}{P^+}\left[\frac{S_T^\alpha k_T^\beta-k_T^\alpha S_T^\beta}{M}h_T^\perp\right].
\label{hTperp}
\end{align}
In the first trace, we have used the identity
\begin{align}
 k_T^2 \epsilon_T^{\alpha\rho} S_{T\rho} &= k_T^\alpha \epsilon_T^{\rho\sigma} k_{T\rho} S_{T\sigma} +(k_T \cdot S_T)\epsilon_T^{\alpha\rho}k_{T\rho},
 \end{align}
and the following combination
\begin{align}
f_T(x,\bk^2)=f_T^\prime(x,\bk^2)-\frac{k_T^2}{2M^2}f_T^\perp(x,\bk^2)
\end{align}
to obtain $f_T$ from $f_T^\prime$ and $f_T^\perp$.

Within the spectator model, we can insert a complete set of the intermediate states $|P-k\rangle$~\cite{Jakob:1997wg} into the correlator (\ref{Phi}), as shown in the left panel of Fig.~\ref{diagrams}.
In the lowest order, the correlator has the form
\begin{align}
\Phi^{(0)}(x,\bm k_T)=\frac{1}{(2\pi)^3}\,\frac{1}{2(1-x)P^+}\,
\overline{\mathcal{M}}^{(0)}\, \mathcal{M}^{(0)} ,
\label{eq:Phi-tree-spect}
\end{align}
where $\mathcal{M}^{(0)}$ is the nucleon-quark-spectator scattering amplitude at the tree level
\begin{align}
\mathcal{M}^{(0)}
&=\langle P-k |\psi(0) |P\rangle \nonumber\\
&=\begin{cases}
  \displaystyle{\frac{i}{k\ssh-m}}\, \Upsilon_s \, U(P)\\
  \displaystyle{\frac{i}{k\ssh-m}}\, \varepsilon^*_{\mu}(P-k,\lambda_a)\,
        \Upsilon_v^{\mu}\, U(P),
  \end{cases}
\label{eq:m-tree}
\end{align}
and $\bar{\mathcal{M}}^{(0)}=(\mathcal{M}^{(0)})^\dag \gamma^0$ corresponds to its Hermitian conjugation,  $\varepsilon_{\mu}(P-k,\lambda_a)$ is the polarization vector of the axial-vector diquark.
The nucleon-quark-diquark vertices  $\Upsilon_{s}$ and $\Upsilon_{v}^\mu$ ($s$ for the scalar diquark and $v$ for the axial-vector diquark) have the form~\cite{Jakob:1997wg}
\begin{align}
\Upsilon_s(k^2) = g_s(k^2),~~~
\Upsilon_v^\mu(k^2)={g_v(k^2)\over \sqrt{2}}\gamma^\mu\gamma^5, \label{eq:vertex}
\end{align}
with $g_X(k^2)$ ($X=s,v$) the form factor for the nucleon-quark-diquark couplings.
In the calculation of T-odd twist-3 TMD distributions, one will encounter the light-cone divergences~\cite{Gamberg:2006ru} when using a point-like coupling.
To regularize these divergences, we choose the dipolar form factor for $g_X(k^2)$:
\begin{align}
g_X(k^2)&= N_X {k^2-m^2\over |k^2-\Lambda_X^2|^2} \nonumber\\
&= N_X{(k^2-m^2)(1-x)^2\over
(\bm k_T^2+L_X^2)^2},~~ X=s,v. \label{eq:gx}
 \end{align}
Here, $N_X$ and $\Lambda_X$ are the normalization constant and the cut-off parameter, respectively, and
$L_X^2$ has the form
\begin{align}
L_X^2=(1-x)\Lambda_{X}^2 +x M_{X}^2-x(1-x)M^2.
\end{align}

Inserting Eqs.~(\ref{eq:m-tree}), (\ref{eq:vertex}) and (\ref{eq:gx}) into Eq.~(\ref{eq:Phi-tree-spect}), we obtain the lowest-order correlator contributed by the scalar diquark component:
\begin{align}
\Phi^{(0)}_s(x,\bk)&\equiv \frac{N_s^2(1-x)^3}{32 \pi^3 P^+}\frac{\left[ (k\ssh +m)\gamma_5 S\ssh (P\ssh +M) (k\ssh +m)\right]}{(\bk^2+L_s^2)^4}, \label{lophis}
\end{align}
and by the axial-vector diquark component:
\begin{align}
 \Phi^{(0)}_{v}(x,\bk)&\equiv \frac{N_v^2(1-x)^3}{64 \pi^3 P^+}d_{\mu\nu}(P-k)\nonumber\\
&\times \frac{\left[(k\ssh +m)\gamma^{\mu}\gamma_5 S\ssh(M-P\ssh )\gamma^{\nu}(k\ssh+m)\right]}{(\bk^2+L_v^2)^4},
\label{lophiv}
\end{align}
where $k^+ = xP^+$ and $d_{\mu\nu}(P-k)=\Sigma_{\lambda} \varepsilon^{*}_{\mu}(\lambda)\varepsilon_{\nu}(\lambda)$ represents  the summation over the polarizations of the axial-vector diquark.

In the calculation of the T-even distributions $h_T(x,\bk^2)$ and $h_T^\perp(x,\bk^2)$, it is sufficient to apply the lowest-order results (\ref{lophis}) and (\ref{lophiv}) of the correlator.
However, to obtain nonzero results for the T-odd distributions $f_T(x,\bk^2)$ and $f_T^\perp (x,\bk^2)$,  one has to consider the nontrivial effect of the gange-link~\cite{Brodsky:2002cx,jy02,Collins:2002plb}, that is, the final-state interaction between the struck quark and the spectator-diquark.
Here we consider the one gluon-exchange approximation on the gauge-link, as shown in the right panel of Fig.~\ref{diagrams}.
Thus the interference of the lowest-order amplitude $\mathcal M^{(0)}$ and the one-loop-order amplitude $\mathcal M^{(1)}$ will give rise to the contribution to the correlator:
\begin{align}
\Phi^{(1)}(x,\bm k_T)=\frac{1}{(2\pi)^3}\,\frac{1}{2(1-x)P^+}\,
\overline{\mathcal{M}}^{(0)}\, \mathcal{M}^{(1)} +h.c..
\label{eq:Phi-spect-loop1}
\end{align}
After some algebra, we arrive at the contributions of the scalar and the arxial-vector diquarks to the correlator at one-loop level:
\begin{align}
  \Phi_s^{(1)}
(x,\bm k_T)
&\equiv
-i e_q N_{s}^2 {(1-x)^2\over 64\pi^3 (P^+)^2}\frac{-i\Gamma^{+}_s}{(\bm{k}_T^2+L_s^2)^2}\nonumber \\
\hspace{-1cm}&\times \int {d^2 \bm q_T\over (2\pi)^2}
{ \left[(\kslash -q\ssh+m)\gamma_5 S\ssh (\Pslash+M)(\kslash +m)\right]
\over \bm q_{T}^2  \left[(\bm{k}_T-\bm{q}_T)^2+L_s^2\right]^2}
 , \label{phis1}\\
 \Phi^{(1)}_{v}
(x,\bm k_T)
&\equiv
-i e_q N_v^2 {(1-x)^2\over 128\pi^3 (P^+)^2}{1\over (\bm{k}_T^2+L_v^2)^2}\nonumber\\
&\times\int {d^2 \bm q_T\over (2\pi)^2} \,
 d_{\rho\alpha}(P-k)\, (-i\Gamma^{+,\alpha\beta}) \nonumber\\
 &\times d_{\sigma\beta}(P-k+q) \nonumber\\
&\times{ \left[(\kslash -q\ssh+m) \gamma^\sigma \gamma_5 S\ssh(M-\Pslash)\gamma^\rho (\kslash +m)\right]
\over \bm q_T^2  \left[(\bm{k}_T-\bm{q}_T)^2+L_v^2\right]^2},
\label{phia1}
\end{align}
with $q^+=0$. Here, $\Gamma_s^\mu $ or $\Gamma_v^{\mu,\alpha\beta}$ is
the vertex between the gluon and the scalar diquark or the axial-vector diquark:
\begin{align}
 \Gamma_s^\mu &= ie_s (2P-2k+q)^\mu, \\
 \Gamma_v^{\mu,\alpha\beta} &=  -i e_v [(2P-2k+q)^\mu g^{\alpha\beta}-(P-k+q)^{\alpha}g^{\mu\beta}\nonumber\\
 &-(P-k)^\beta g^{\mu\alpha}]\label{Gamma},
\end{align}
where $e_{s/v}$ denotes the charge of the scalar/axial-vector diquark.

Substituting (\ref{lophis}) into (\ref{hT}) and (\ref{hTperp}), we obtain the T-even  distributions $h_T$ and $h_T^\perp$ contributed by the scalar diquark:
\begin{align}
h_T^s(x,\bk^2)&=\frac{{N_s}^2(1-x)^2}{16\pi^3}
\frac{\left[(1-x)^2M^2-\bk^2-M_s^2\right]}{(\bk^2+L_s^2)^4},\\
h_T^{\perp s}(x,\bk^2)&=\frac{N_s^2(1-x)^2}{16\pi^3}\frac{1}{(\bk^2+L_s^2)^4}\nonumber\\
&\times\left[(1-x)(M^2+2mM+xM^2)-\bk^2-M_s^2\right].
\end{align}
We find that the above expressions are in consistence with the results in Ref.~\cite{Jakob:1997wg}.
Similarly, we get the scalar diquark contributions to the T-odd distributions $f_T$ and $f_T^\perp$:
\begin{align}
f_T^s(x,\bk^2)&=-\frac{{N_s}^2(1-x)^2}{32\pi^3} \frac{e_s e_q}{4\pi}\frac{(x+\frac{m}{M})(L_s^2-\bk^2)}{L_s^2(L_s^2+\bk^2)^3},\label{eq:fts}\\
f_T^{\perp s}(x,\bk^2)&=0,
\end{align}
which have already been presented in Ref.~\cite{Lu:2012gu}.

\begin{figure}
  % Requires \usepackage{graphicx}
  \includegraphics[width=0.49\columnwidth]{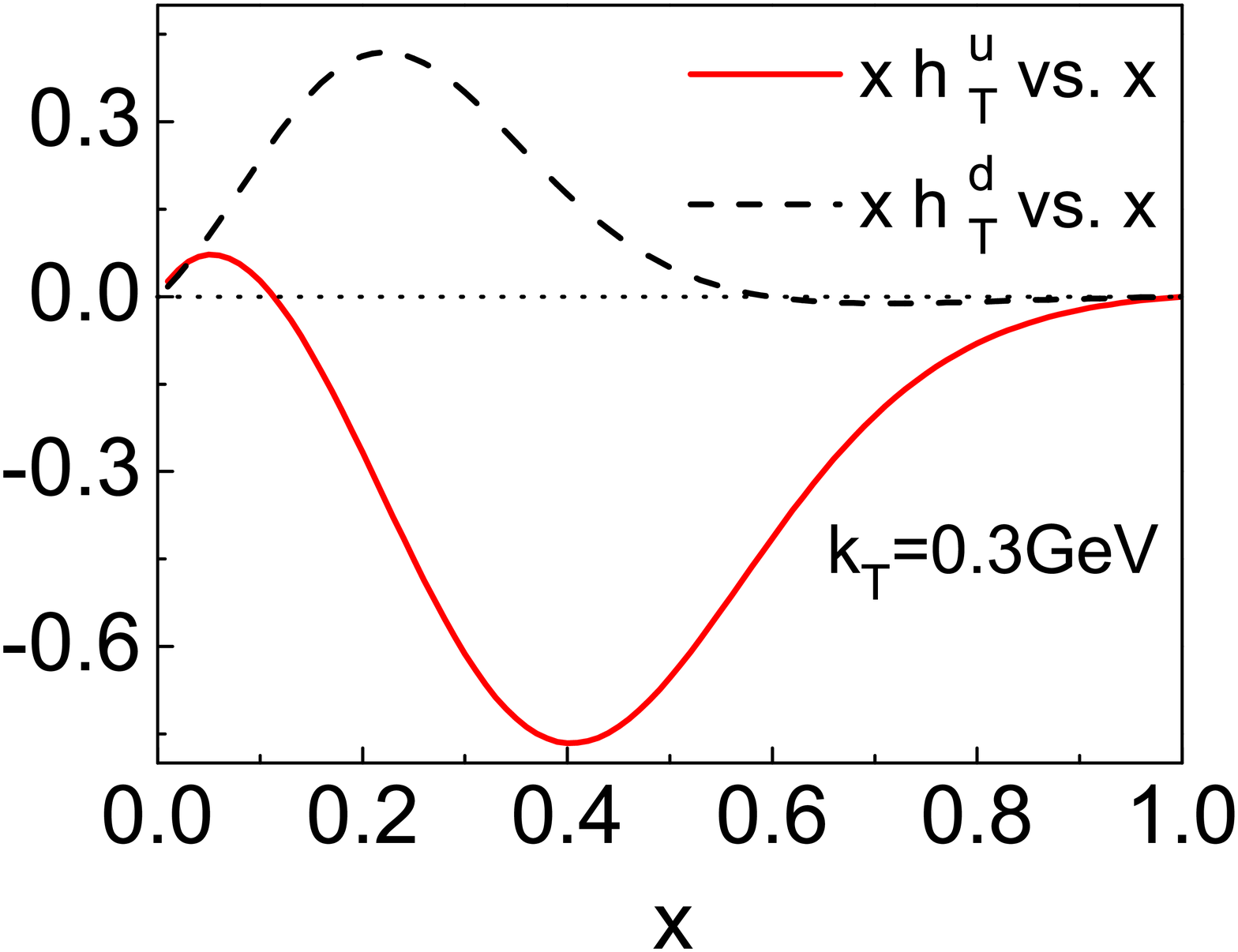}
  \includegraphics[width=0.49\columnwidth]{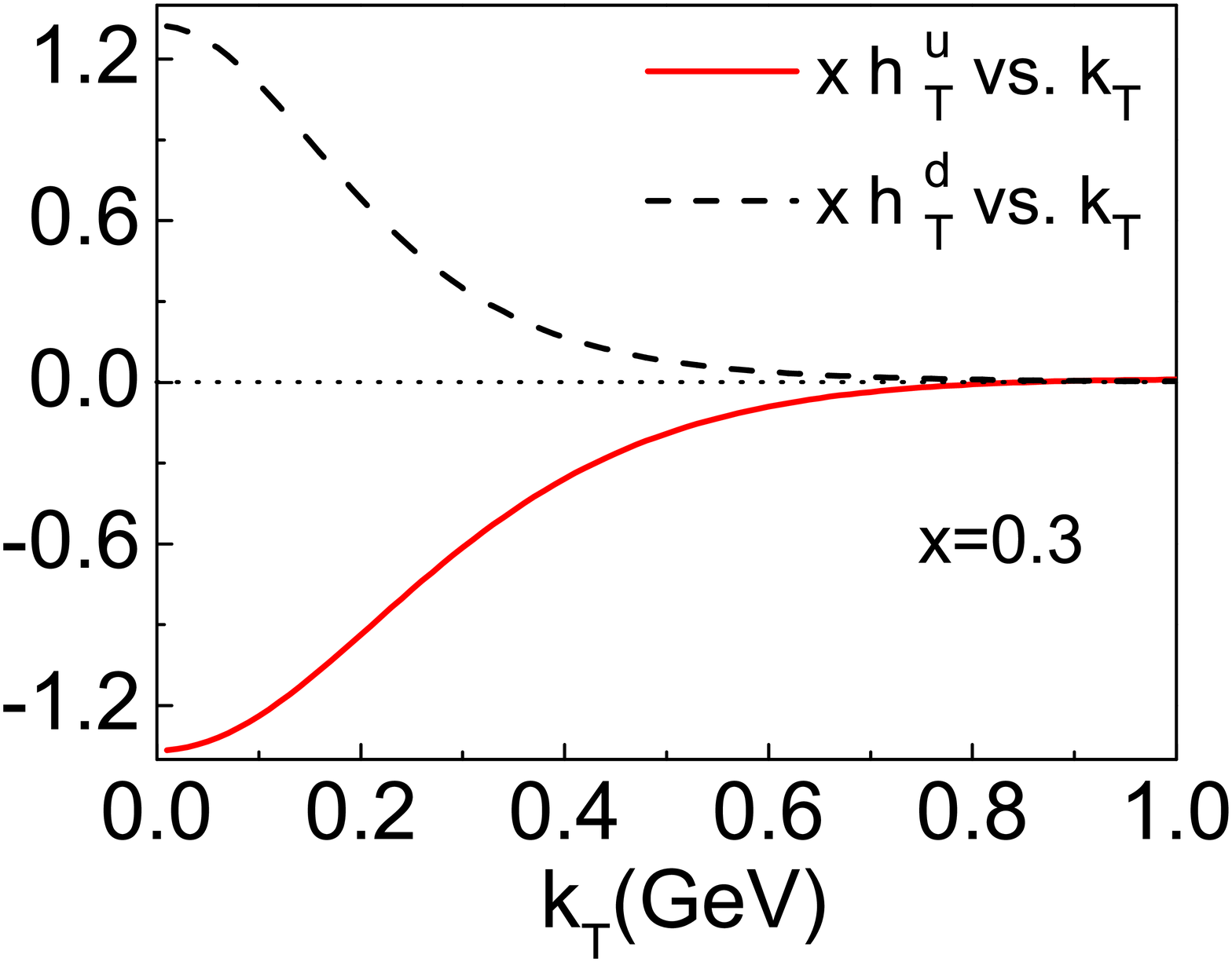}
  \caption{Left panel: model results for $x h_T^{u}$ (solid line) and $x h_T^{d}$ (dashed line) as functions of $x$ at $k_T=0.3\,\text{GeV}$; right panel: model results for $x h_T^{u}$ (solid line) and $x h_T^{d}$ (dashed line) as functions of $k_T$ at $x=0.3$.}\label{FIG:ht}
\end{figure}
\begin{figure}
  % Requires \usepackage{graphicx}
  \includegraphics[width=0.49\columnwidth]{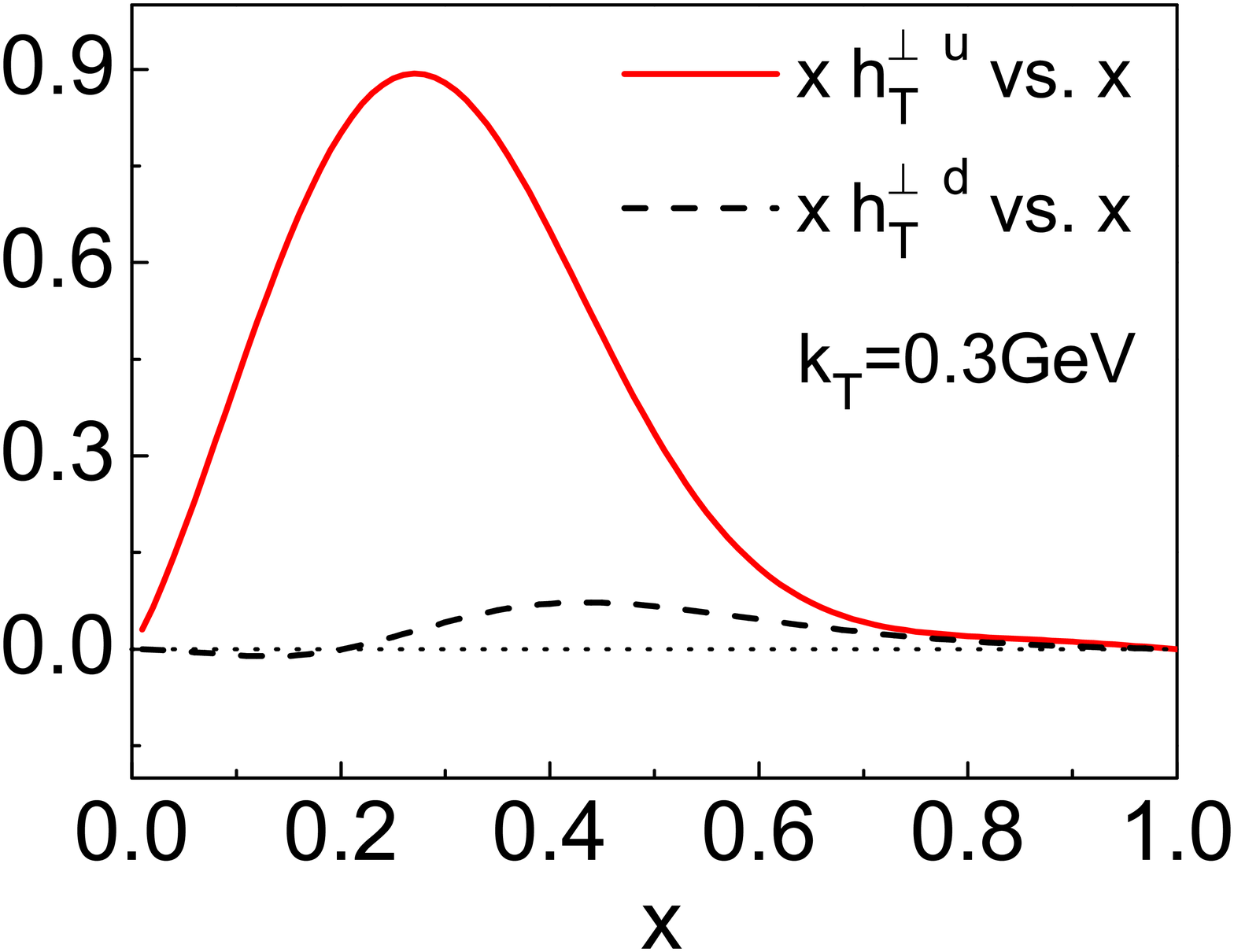}
  \includegraphics[width=0.49\columnwidth]{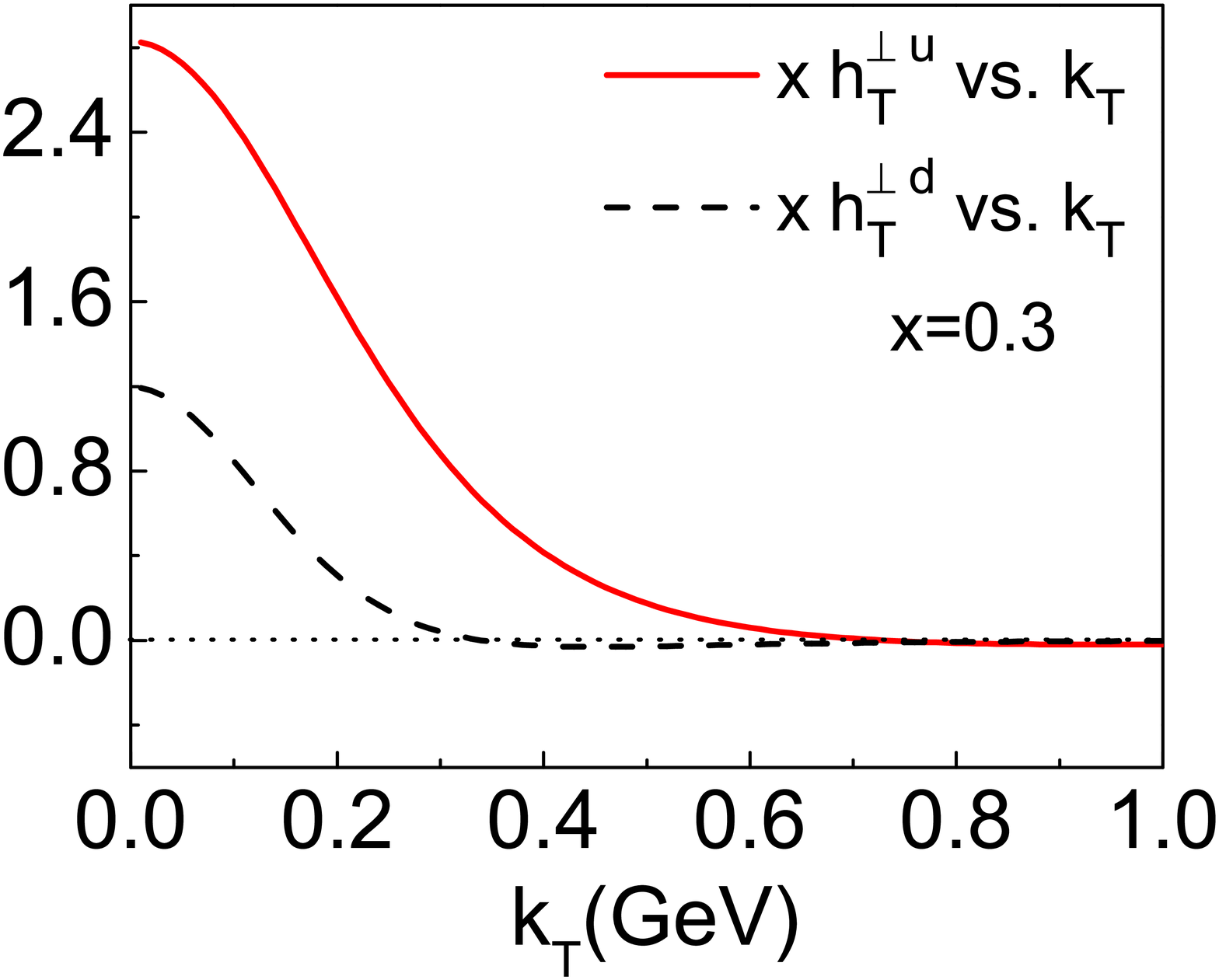}
  \caption{Similar to Fig.~\ref{FIG:ht}, but for the model results of $x h_T^{\perp u}$ (solid line) and $x h_T^{\perp d}$ (dashed line).}\label{FIG:htperp}
\end{figure}

To calculate the quark correlator contributed by the axial-vector diquark, we adopt the following form for the propagator $d_{\mu\nu}$:
\begin{align}
 d_{\mu\nu}(P-k)  =& \,-g_{\mu\nu}\,+\, {(P-k)_\mu n_{-\nu}
 \,+ \,(P-k)_\nu n_{-\mu}\over(P-k)\cdot n_-}\,\nonumber\\
 & - \,{M_v^2 \over\left[(P-k)\cdot n_-\right]^2 }\,n_{-\mu} n_{-\nu} ,\label{d1}
\end{align}
which is the summation over the light-cone transverse polarizations of the axial-vector diquark~\cite{Brodsky:2000ii}.
This form has been applied to calculate leading-twist TMD distributions in Ref.~\cite{Bacchetta:2008af}.
Similar to the scalar-diquark case, we obtain the distributions contributed by the axial-vector diquark component:
\begin{align}
h_T^v(x,\bk^2)&=\frac{N_v^2(1-x)}{16\pi^3}\frac{1}{(\bk^2+L_v^2)^4}\nonumber\\
&\times\left[(1-x)(m^2+2xmM+xM^2)\right.\nonumber\\
&\left.+\bk^2-xM_v^2\right],
\label{hTa}\\
h_T^{\perp v}(x,\bk^2)&=\frac{N_v^2(1-x)}{16\pi^3}\frac{1}{(\bk^2+L_v^2)^4}\nonumber\\
&\times\left[(1-x)(m^2-xM^2)-\bk^2+xM_v^2\right],
\label{hTperpa}\\
f_T^v(x,\bk^2)&=0,
\label{fTa}\\
f_T^{\perp v}(x,\bk^2)&=-\frac{N_v^2(1-x)^2M(m+xM)}{16\pi^3(L_v^2+\bk^2)^2\bk^2}\frac{e_v e_q}{4\pi}\nonumber\\
&\times\left[\frac{1}{\bk^2}\ln{\frac{\bk^2+L_v^2}{L_v^2}}+\frac{\bk^2-L_v^2}{L_v^2 (L_v^2+\bk^2)}\right].
\label{fTperpa}
\end{align}

The distributions for the $u$ and $d$ valence quarks can be constructed from $f^{s}$ and $f^{v}$ obtained previously.
Here we follow the approach in Ref.~\cite{Bacchetta:2008af}, in which
the two isospin states (isoscalar and isovector) of the axial-vector diquark are distinguished:
\begin{align}
f^u=c_s^2 f^s + c_a^2 f^a,~~~~f^d=c_{a^\prime}^2 f^{a^\prime}\label{ud},
\end{align}
where $c_s$, $c_a$ and $c_{a^\prime}$ are the free parameters of the model, $a$ and $a^\prime$ denote the isoscalar and isovector states of the axial-diquark, respectively.
These parameters, together with the mass parameters (such as the diquark masses $M_X$, the cut-off parameters $\Lambda_X$), are fitted from the ZEUS~\cite{Chekanov:2002pv} and GRSV01~\cite{Gluck:2000dy} parton distributions.
Finally, we use the following replacement for the combination of the charges of the quark $q$ and the spectator diquark $X$ to convert our calculation to that in QCD
\begin{align}
{e_qe_X\over 4 \pi}\rightarrow -  C_F \alpha_s.
\end{align}
In this work, the coupling constant $\alpha_s$ is chosen as $\alpha_s \approx 0.3 $.

\begin{figure}
  % Requires \usepackage{graphicx}
  \includegraphics[width=0.49\columnwidth]{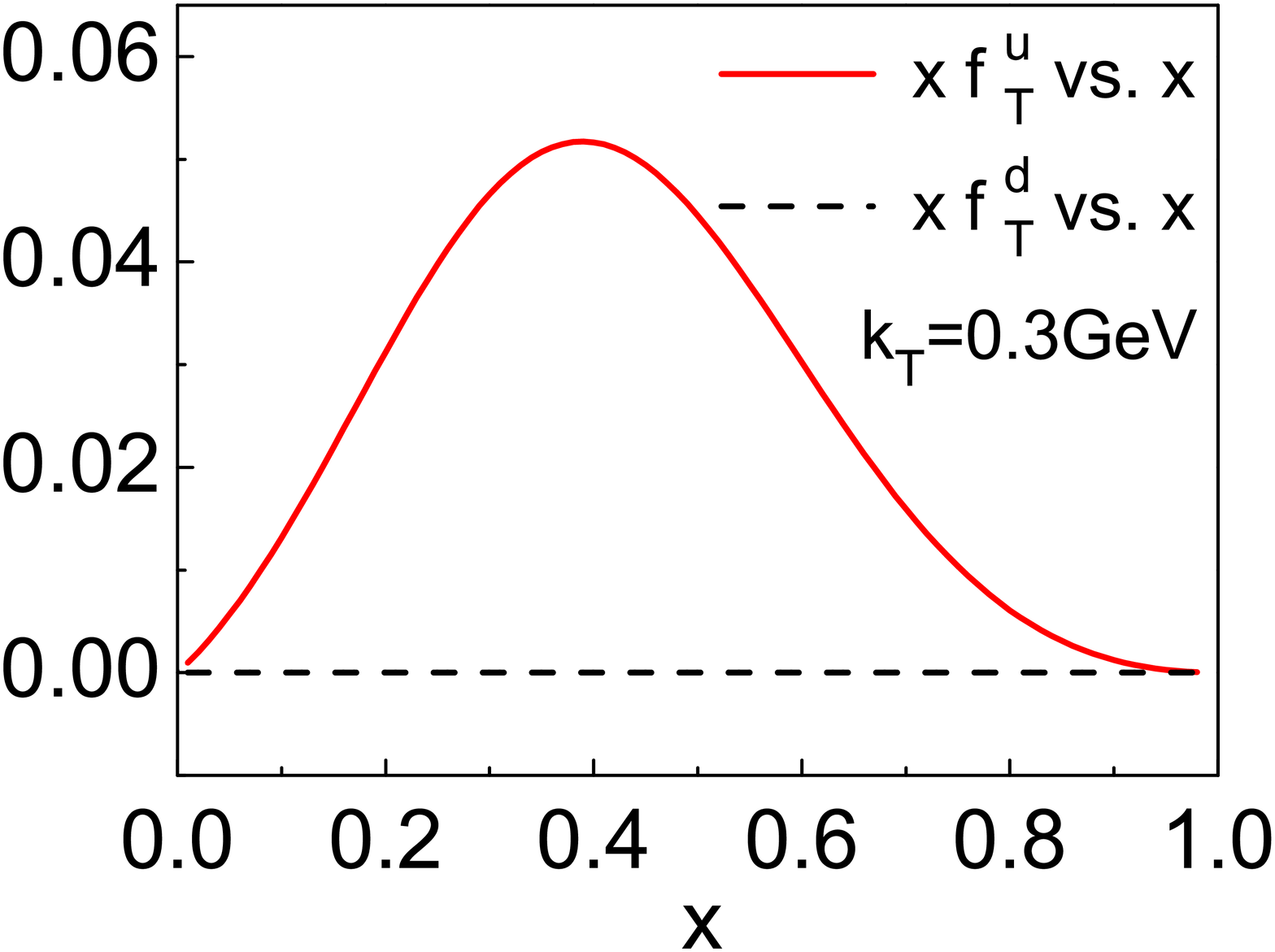}
  \includegraphics[width=0.49\columnwidth]{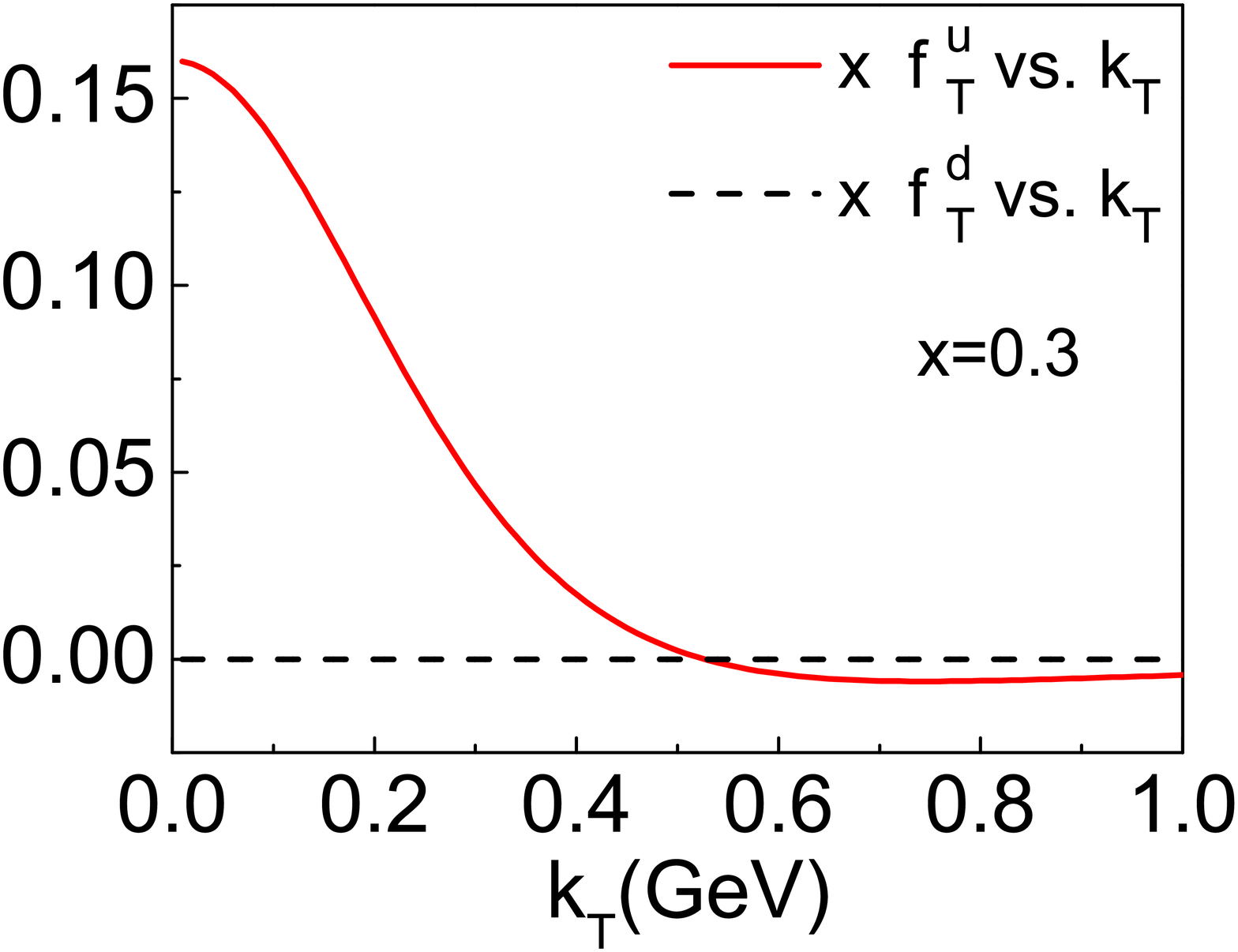}
  \caption{Similar to Fig.~\ref{FIG:ht}, but for the model results of $x f_T^{u}$ (solid line) and $x f_T^{d}$ (dashed line).}\label{FIG:ft}
\end{figure}
\begin{figure}
  % Requires \usepackage{graphicx}
  \includegraphics[width=0.49\columnwidth]{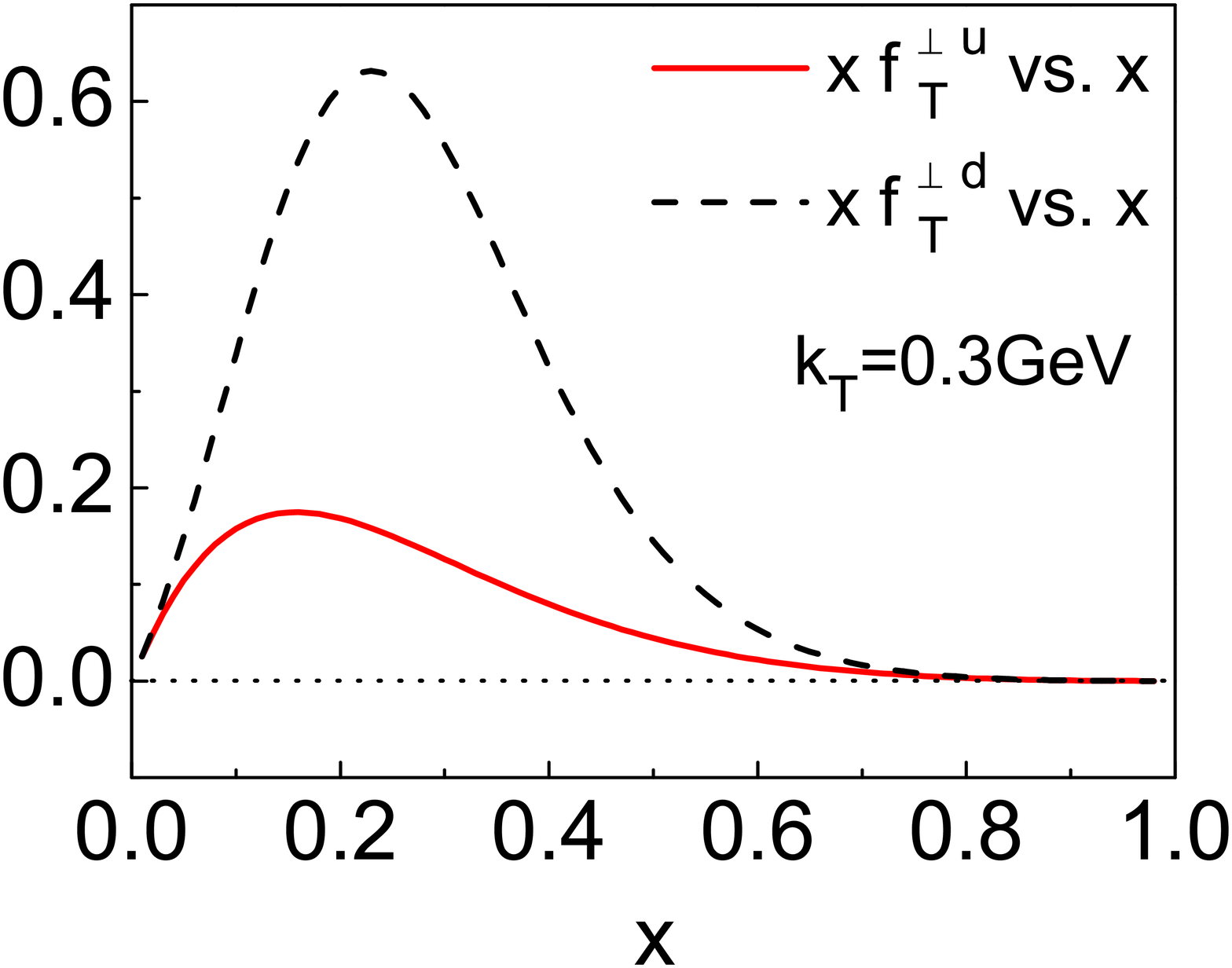}
  \includegraphics[width=0.49\columnwidth]{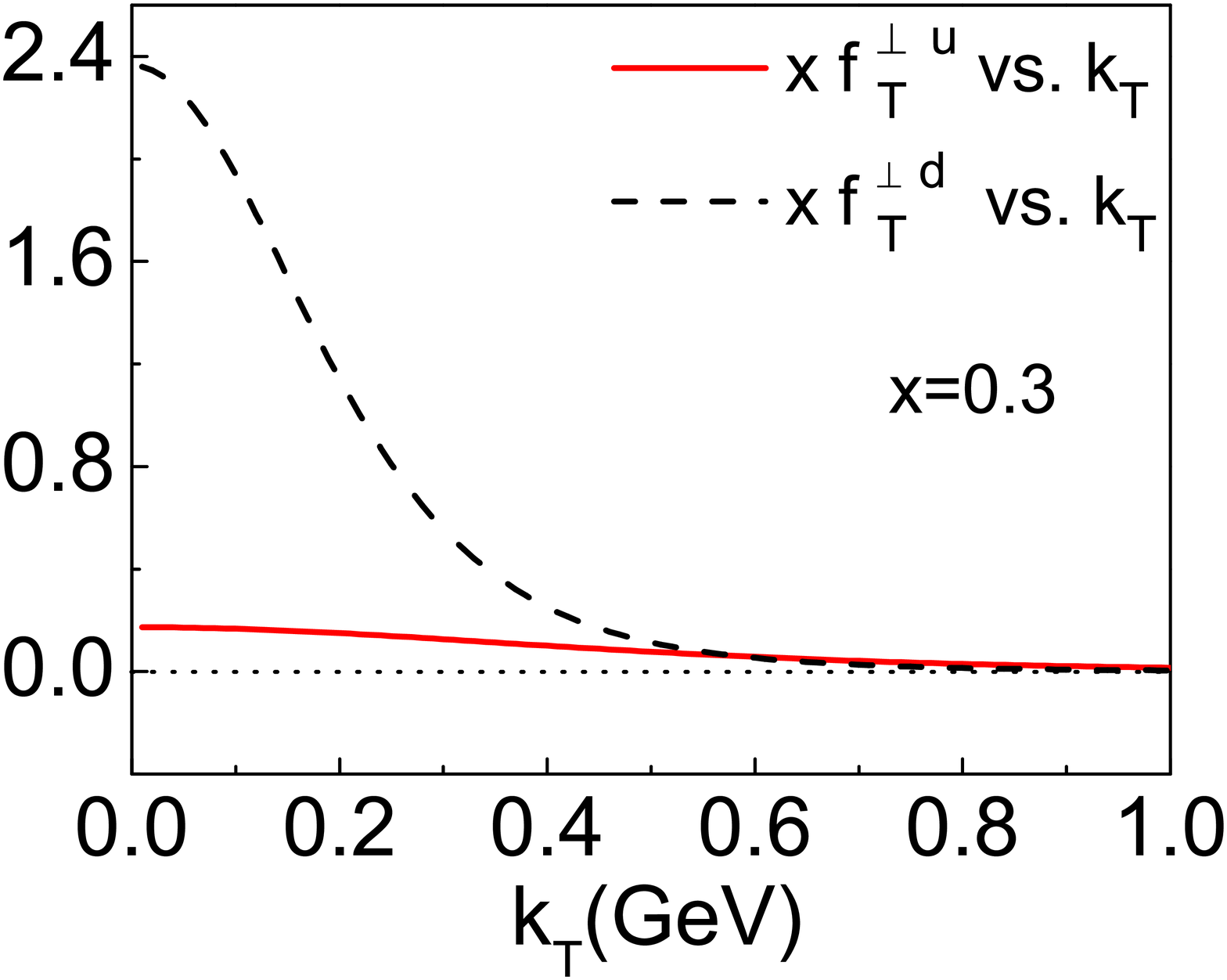}
  \caption{Similar to Fig.~\ref{FIG:ht}, but for the model results of $x f_T^{\perp u}$ (solid line) and $x f_T^{\perp d}$ (dashed line).}\label{FIG:ftperp}
\end{figure}

In Fig.~\ref{FIG:ht}, we plot the distribution $h_T$ as functions of $x$ (left panel) and $k_T$ (right panel), respectively.
The solid and dashed curves show the results for the $u$ and $d$ valence quarks, respectively.
As we can see, in the specified kinematic region ($x=0.3$ or $k_T=0.3$ GeV), the distributions $h_T^u$ and $h_T^d$ have similar sizes but opposite signs.
Similarly, we present the plots of the distribution $h_T^\perp$ in Fig.~\ref{FIG:htperp}, which shows that $h_T^{\perp\,u}$ is positive and its size is much larger than that of $h_T^{\perp\,d}$.

In Figs.~\ref{FIG:ft}, we show the curves of the T-odd distribution $f_T$.
Since $f_T^v$ vanishes in the model we adopt, here $f_T^d$ is zero.
We find that at low $k_T$, $f_T^u$ is positive, while it is negative in the intermediate range of $k_T$, and eventually falls to zero at large $k_T$. That is, there is a node of the distribution $f_T^u$ in $k_T$.
The size of $f_T$ is much smaller compared to those of the T-even distributions $h_T$ and $h_T^\perp$.
Specially, with the $k_T$-dependence of $f_T$ given in Eq.~(\ref{eq:fts}), we can verify that $f_T^u$ vanishes when it is integrated over the transverse momentum ~\cite{Goeke:2005hb}:
\begin{align}
\int d^2\bk f_T^u(x,\bk^2)=0.
\end{align}
This is an expected result from the time-reversal invariance for integrated distributions, and it indicates that the distribution $f_T$ will not give any contribution to the transverse SSA in inclusive DIS process~\cite{Airapetian:2009ab,Metz:2012ui}.

Finally, in Fig.~\ref{FIG:ftperp}, we plot the second T-odd distribution $f_T^\perp$.
The results show that $f_T^{\perp d}$ dominates over $f_T^{\perp u}$ in the chosen kinematic regime.
This may be explained by the fact that $f_T^{\perp s}$ is zero and only $f_T^{\perp v}$ contributes in our model.

\section{Prediction on the transverse SSAs for charged and neutral pions in SIDIS}
\label{BSAs}

In this section, we perform our predictions on the transverse SSAs at twist-3 level
in SIDIS:
\begin{align}
l (\ell) \, + \, p^\uparrow (P) \, \rightarrow \, l' (\ell')
\, + \, h (P_h) \, + \, X (P_X)\,,
\label{sidis}
\end{align}
where $\uparrow$ denotes the transverse polarization of the proton target, $\ell$ and $\ell'$ represent the momenta of the incoming and outgoing leptons, and $P$ and $P_h$ denote the momenta of the target nucleon and the final-state hadron.

\begin{figure}
 % Requires \usepackage{graphicx}
  \includegraphics[width=0.8\columnwidth]{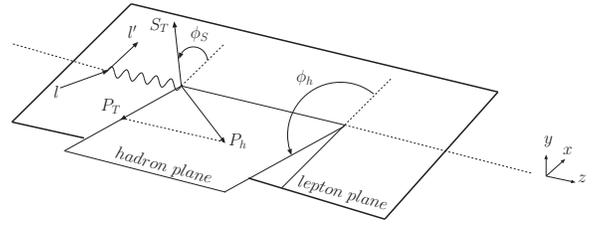}
 \caption {The kinematical configuration for the polarized SIDIS process. The initial and scattered leptonic momenta define the lepton plane ($x-z$ plane), while the detected hadron momentum together with the $z$ axis identify the hadron production plane.}
 \label{SIDISframe}
\end{figure}

Following the Trento convention~\cite{Bacchetta:2004jz}, in our calculation we adopt the reference frame shown in Fig.~\ref{SIDISframe}, where $\bm P_T$ and $\bm {S}_T$ are the transverse momentum of the detected pion and the transverse spin of the target, and their azimuthal angles with respect to the lepton plane are denoted by $\phi_h$ and $\phi_S$.
The invariant variables used to express the differential cross section of SIDIS are defined as
\begin{align}
&x = \frac{Q^2}{2\,P\cdot q},~~~
y = \frac{P \cdot q}{P \cdot l},~~~
z = \frac{P \cdot P_h}{P\cdot q},~~~
\gamma={2M x\over Q},~~~\nonumber\\
&Q^2=-q^2, ~~~
s=(P+\ell)^2,~~~
W^2=(P+q)^2,~~~
\end{align}
where $q=\ell-\ell'$ is the four-momentum of the virtual photon, and $W$ is the invariant mass of the hadronic final state.

With the above variables, the differential cross section of the process (\ref{sidis}) for an unpolarized beam scattering off a transversely polarized hadron can be expressed as~\cite{Bacchetta:2006tn}
\begin{align}
\frac{d\sigma}{d\xbj dy\,dz d\phi_S d\ph dP^2_T} &=\frac{\alpha^2}{\xbj y Q^2}\frac{y^2}{2(1-\varepsilon)}
 \Bigl( 1+ \frac{\gamma^2}{2\xbj} \Bigr)
  \left\{F_{\mathrm{UU}} \right.\nonumber\\
  &+ \left.|\bm{S}_T| \left[\sqrt{2\,\varepsilon (1+\varepsilon)}\,\left(\sin\phi_S\,F_{\mathrm{UT}}^{\sin \phi_S }\right.\right.\right.\nonumber\\
&+ \left.\left.
 \left. \sin(2\phi_h-\phi_S)\,
F_{\mathrm{UT}}^{\sin{(2\phi_h -\phi_S)}}
\right)\right]+\cdots\right\}.\label{FUT}
\end{align}
Here, $F_{\mathrm{UU}}$ is the spin-averaged structure function, and $F_{\mathrm{UT}}^{\sin{\phi_S}}$ and $F_{\mathrm{UT}}^{\sin{(2\phi_h -\phi_S)}}$ are the spin-dependent structure functions that contribute to the $\sin\phi_S$ and $\sin(2\phi_h-\phi_S)$ azimuthal asymmetries, respectively.
The ellipsis stands for the other three spin-dependent structure functions $F_{\mathrm{UT}}^{\sin{\phi_h - \phi_S}}$, $F_{\mathrm{UT}}^{\sin{\phi_h + \phi_S}}$ and $F_{\mathrm{UT}}^{\sin{(3\phi_h -\phi_S)}}$, which are separately contributed by the $f_{1T}^\perp D_1$, $h_1 H_1^\perp$ and $h_{1T}^\perp H_1^\perp$ terms, and which will not be studied in this work.
The ratio of the longitudinal and transverse photon flux $\varepsilon$ is given as
\begin{align}
\varepsilon=\frac{1-y-\gamma^2y^2/4}{1-y+y^2/2+\gamma^2y^2/4}.
\end{align}

Based on the tree-level factorization adopted in Ref.~\cite{Bacchetta:2006tn}, in the parton model, the structure functions in Eq.~(\ref{FUT}) can be expressed as the convolutions of twist-2 and twist-3 TMD distributions and FFs. With the notation
\begin{align}
\mathcal{C}[w fD] &=x\sum_q e_q^2\int d^2\bm k_T\int d^2 \bm p_T\delta^2(z\bm k_T-\bm P_T+\bm p_T) \nonumber\\
&\times w(\bm k_T, \bm p_T)f^q(x,\bm k_T^2) D^q(z,\bm p_T^2),
\end{align}
$F_{\mathrm{UU}}$, $F_\mathrm{{UT}}^{\sin{\phi_S}}$ and $F_{\mathrm{UT}}^{\sin{(2\phi_h -\phi_S)}}$ may be given as~\cite{Bacchetta:2006tn}
\begin{align}
F_{\mathrm{UU}} &= \mathcal{C}[f_1 D_1], \label{FUU}\\
F_{\mathrm{UT}}^{\sin\phi_S}&\approx\frac{2M}{Q}{\cal C}\left\{x f_T D_1\right.\nonumber\\
&+\left.\frac{\bp \cdot \bk}{2z M M_h}\left(x h_T H_1^\perp-x h_T^\perp H_1^\perp\right)
\right\},
\label{FUT1} \\
F_{\mathrm{UT}}^{\sin{(2\phi_h-\phi_S)}}&\approx\frac{2M}{Q}{\cal C}\left\{\frac{2(\hat{\bP}\cdot \bk)^2-\bk^2}{2M^2}\left(x f_T^{\perp}D_1\right)\right.\nonumber\\
&+\frac{2(\hat{\bP}\cdot \bp)(\hat{\bP}\cdot\bk)-\bp\cdot \bk}{2zM M_h}\nonumber\\
&\times\left.\left[x h_T H_1^\perp+x h_T^\perp H_1^\perp\right]\right\}.
\label{FUT2}
\end{align}
Here, $M_h$ is the mass of the final-state hadron and $\hat {\bm P}_T = {\bP \over P_T}$ with $P_T =|\bP|$.
As stated in Section I, we have neglected the contributions from the twist-3 TMD FFs $\tilde{D}^\perp$, $\tilde{G}^\perp$ and $\tilde{H}$ in Eqs.~(\ref{FUT1}) and (\ref{FUT2}).
Therefore, with Eqs.~(\ref{FUU}), (\ref{FUT1}) and (\ref{FUT2}), the $P_T$-dependent transverse SSAs $A_{\mathrm{UT}}^{\sin\phi_S}$ and $A_{\mathrm{UT}}^{\sin{(2\phi_h-\phi_S)}}$ can be defined as
\begin{align}
A_{\mathrm{UT}}^{\sin\phi_S}(P_T) &= \frac{\int dx \int dy \int dz \;\mathcal{C}_{\mathrm{UT}}\;F_{\mathrm{UT}}^{\sin{\phi_S}}}{\int dx \int dy \int dz \;\mathcal{C}_{\mathrm{UU}}\;F_{\mathrm{UU}} }\;,\label{asy1}
\end{align}
\begin{align}
A_{\mathrm{UT}}^{\sin{(2\phi_h-\phi_S)}}(P_T) &= \frac{\int dx \int dy \int dz \;\mathcal{C}_{\mathrm{UT}} \;F_{\mathrm{UT}}^{\sin{(2\phi_h-\phi_S)}}}{\int dx \int dy \int dz \;\mathcal{C}_{\mathrm{UU}}\;F_{\mathrm{UU}} }\;, \label{asy2}
\end{align}
where we have defined the kinematical factors
\begin{align}
\mathcal{C}_{\mathrm{UU}}&=\frac{1}{x y Q^2}\frac{y^2}{2(1-\varepsilon)}\Bigl( 1+ \frac{\gamma^2}{2x}
\Bigl),\\
\mathcal{C}_{\mathrm{UT}}&=\frac{1}{x y Q^2}\frac{y^2}{2(1-\varepsilon)}\Bigl( 1+ \frac{\gamma^2}{2x} \Bigr) \sqrt{2\varepsilon(1+\varepsilon)}.
\end{align}
The $x$-dependent and the $z$-dependent asymmetries can be defined in a similar way.

In order to give the numerical prediction on transverse SSAs, we need to know the unpolarized TMD distribution $f_1(x,\bm k_T^2)$ and FF $D_1(z,\bm p_T^2)$, as well as the Collins function $H_1^\perp(z,\bm p_T^2)$.
For consistency, we use the same model result~\cite{Bacchetta:2006tn} for $f_1$, which is fitted from the ZEUS~\cite{Chekanov:2002pv} data set on the unpolarized distribution.
For the TMD FF $D_1^q(z,\bp^2)$, we assume its $p_T$ dependence has a Gaussian form
\begin{align}
D_1^q\left(z,\bp^2\right)=D_1^q(z)\, \frac{1}{\pi \langle \bp^2\rangle}
\, e^{-\bm p_T^2/\langle \bp^2\rangle},
\end{align}
where $\langle \bp^2\rangle$ is the Gaussian width for $\bp^2$, and we choose
its value as 0.2 \textrm{GeV}$^2$, following the result in Ref.~\cite{Anselmino:2005nn}.
For the integrated FFs $D_1^q(z)$, we adopt the leading-order set of the DSS parametrization~\cite{deFlorian:2007aj}.
As for the Collins function $H_1^{\perp}$, we adopt the relations below for different pion productions:
\begin{align}
 H_1^{\perp \pi^+/u}&=H_1^{\perp \pi^-/d}\equiv H_{1 fav}^{\perp} ,\\
 H_1^{\perp \pi^+/d}&=H_1^{\perp \pi^-/u}\equiv H_{1 unf}^{\perp} ,\\
 H_1^{\perp \pi^0/u}&=H_1^{\perp \pi^0/d}\equiv{1\over 2}\left( H_{1 fav}^{\perp}+H_{1 unf}^{\perp}\right),
\label{collins}
\end{align}
where $H_{1 fav}^{\perp}$ and $H_{1 unf}^{\perp}$ are the favored and unfavored Collins functions, for which we apply the parameterizations from Ref.~\cite{Anselmino:2008jk}.

\begin{figure}
  \includegraphics[width=0.99\columnwidth]{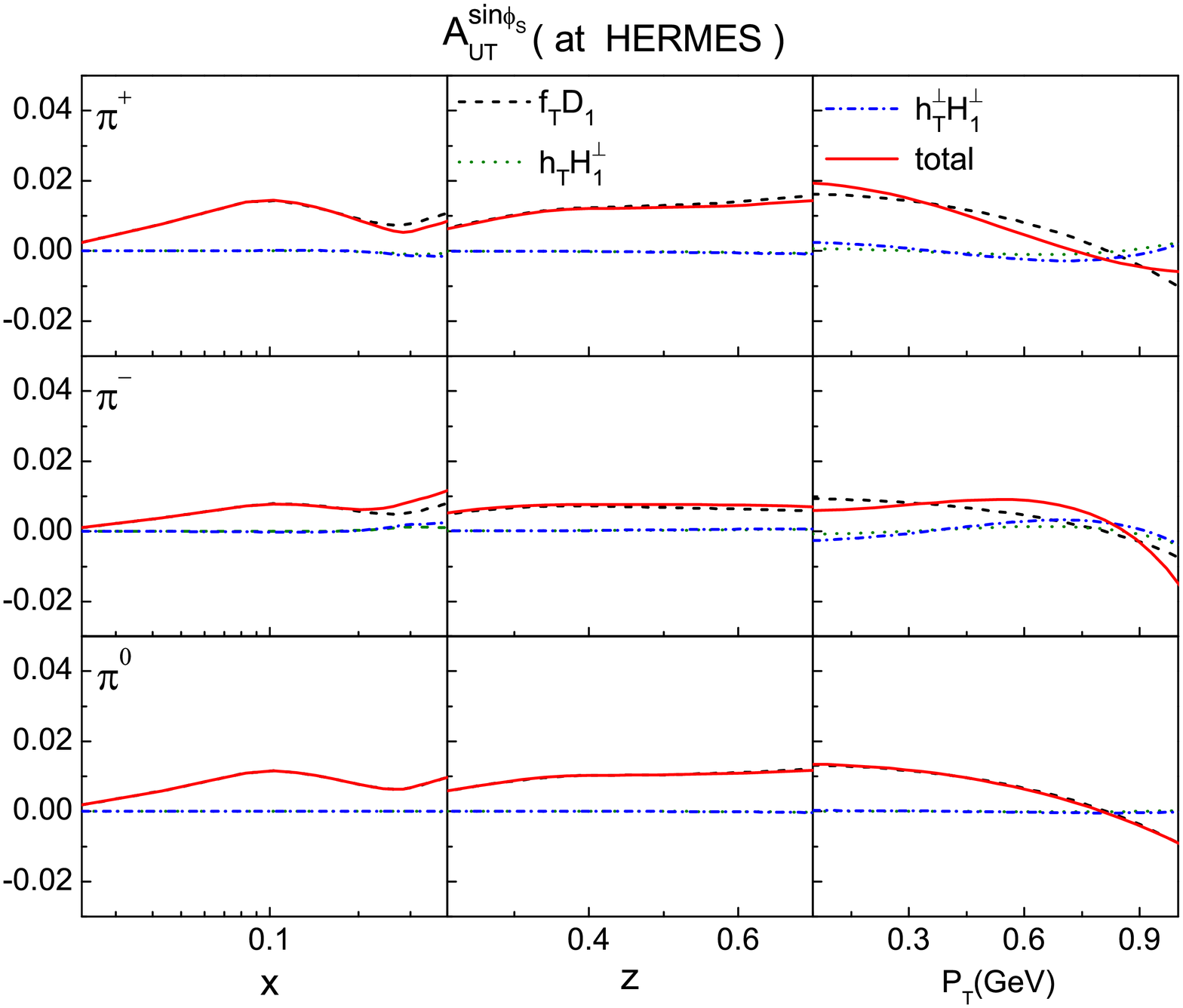}
  \caption{Prediction on the transverse SSA $A_{\mathrm{UT}}^{\sin\phi_S}$ for $\pi^+$ (up panel), $\pi^-$ (middle panel) and $\pi^0$ (down panel) in SIDIS at HERMES. The dashed, dotted and dash-dotted curves represent the asymmetries from the $f_T D_1$, $h_T H_1^\perp$ and $h_T^\perp H_1^\perp$ terms, respectively.  The solid curves correspond to the total contribution.}
  \label{HERMESa}
\end{figure}

Finally, we also consider the following kinematical constraints~\cite{Boglione:2011wm} on the intrinsic transverse momenta of the initial quarks in our calculation:
\begin{equation}
 \begin{cases}
\bm k_{T}^2\leq(2-x)(1-x)Q^2, ~~~\textrm{for}~~0< x< 1 %& (\text{energy bound})
; \\
\bm k_{T}^2\leq \frac{x(1-x)} {(1-2x)^2}\, Q^2, ~~~~~~~~~~~~\textrm{for}~~x< 0.5.
\end{cases}\label{constraints}
 \end{equation}
The former is obtained by requiring the energy of the parton to be less than the energy of the parent hadron; while the later is given by the requirement that the parton should move in the forward direction with respect to the parent hadron~\cite{Boglione:2011wm}.
For the region $x<0.5$, there are two upper limits for $\bm k_T^2$ applied in the region $x<0.5$ at the same time; it is understood that the smaller one should be chosen.

\subsection{HERMES}
\label{prediction}

To perform numerical calculation on the transverse SSAs of charged and neutral pion production in SIDIS at HERMES, which can be performed by using an unpolarized positron beam at the energy of $27.6 \,\textrm{GeV}$ scattered off a transversely polarized proton target, we adopt the following kinematical cuts~\cite{Airapetian:2009ae}:
\begin{align}
&0.023 < x < 0.4,~~0.1 < y < 0.95,~~0.2 < z < 0.7, \nonumber\\
&W^2 > 10\, \textrm{GeV}^2,~~~Q^2 > 1 \textrm{GeV}^2, \nonumber\\
&0.05 < P_T < 1.2\,\textrm{GeV},~~2\,\textrm{GeV} < E_h < 15\, \textrm{GeV},
\end{align}
where $E_h$ is the energy of the detected pion in the target rest frame.

\begin{figure}
  \includegraphics[width=0.99\columnwidth]{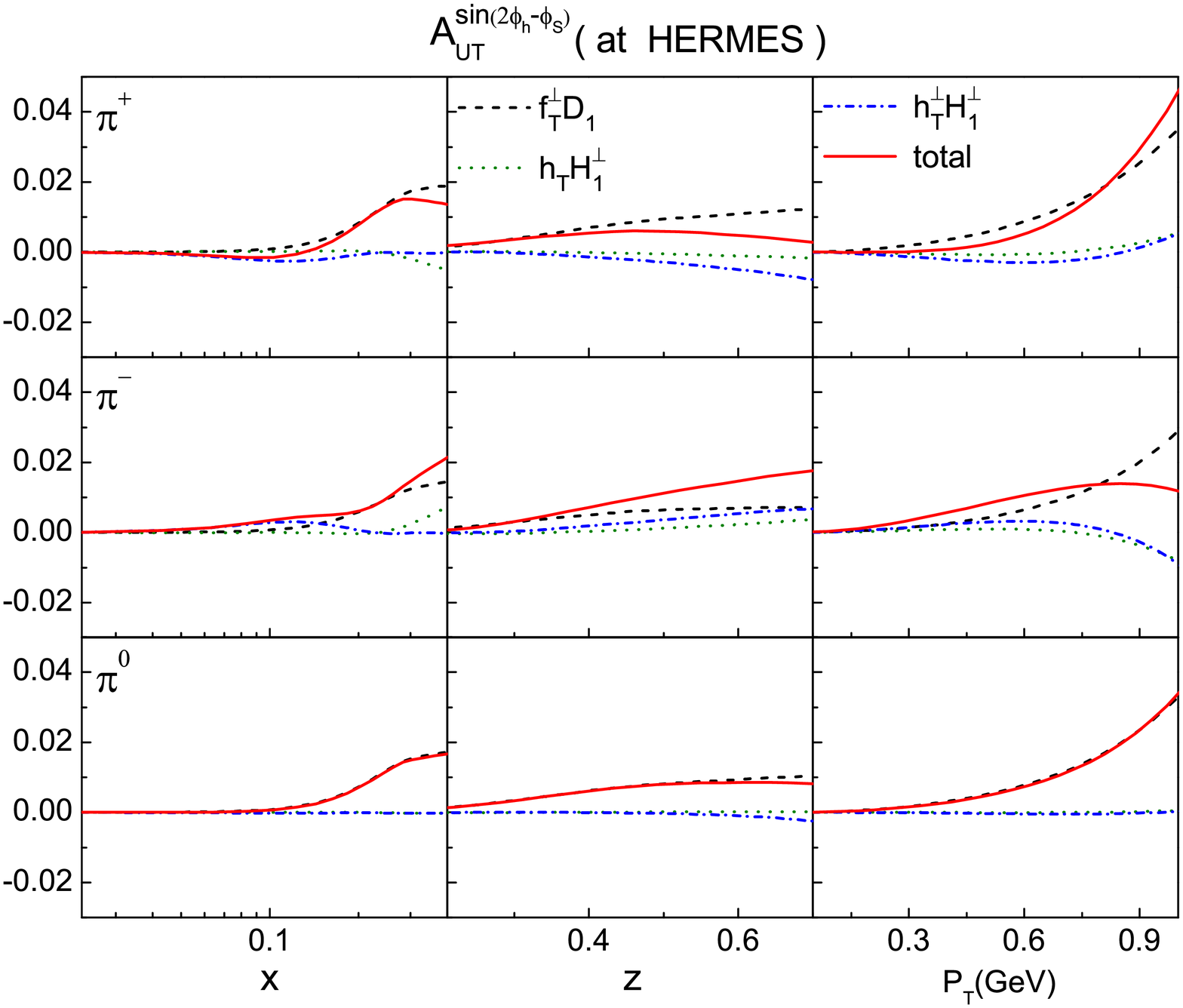}
  \caption{Similar to Fig.~\ref{HERMESa}, but on the asymmetry $A_{\mathrm{UT}}^{\sin(2\phi_h -\phi_S)}$.
  The dashed, dotted and dash-dotted curves show the asymmetries from the $f_T^\perp D_1$, $h_T H_1^\perp$ and $h_T^\perp H_1^\perp$ terms, respectively. The solid curves correspond to the total contribution.}
  \label{HERMESb}
\end{figure}

In the left, central and right panels of Fig.~\ref{HERMESa}, we show our prediction on the transverse SSA $A_{\mathrm{UT}}^{\sin \phi_S}$ at HERMES for $\pi^+$, $\pi^-$ and $\pi^0$ as functions of $x$, $z$, and $P_T$.
To distinguish the origins of different contributions, we use the dashed, dotted and dash-dotted curves to denote the contributions from the $f_T D_1$, $h_T H_1^\perp$ and $h_T^\perp H_1^\perp$ term, respectively.
The solid curves stand for the total contribution.
As we can see, the asymmetry $A_{\mathrm{UT}}^{\sin \phi_S}$ is positive, the size is around $1\%$ to $2\%$ at the kinematics of HERMES, and the dominant contribution is from the $f_T D_1$ term for all three pions.
The contributions from the $h_T H_1^\perp$ and $h_T^\perp H_1^\perp$ terms are nearly negligible except in the larger $P_T$ region for charged pions.
This is due to the kinematical factor $\bp \cdot \bk/(2z M M_h)$ associated with the $h_T H_1^\perp$ and $h_T^\perp H_1^\perp$ terms, and also the fact that the size of $H_1^\perp$ is less than the size of $D_1$, despite the size of $h_T$ or $h_T^\perp$ is larger than that of the T-odd distribution $f_T$.

\begin{figure}
  \includegraphics[width=0.99\columnwidth]{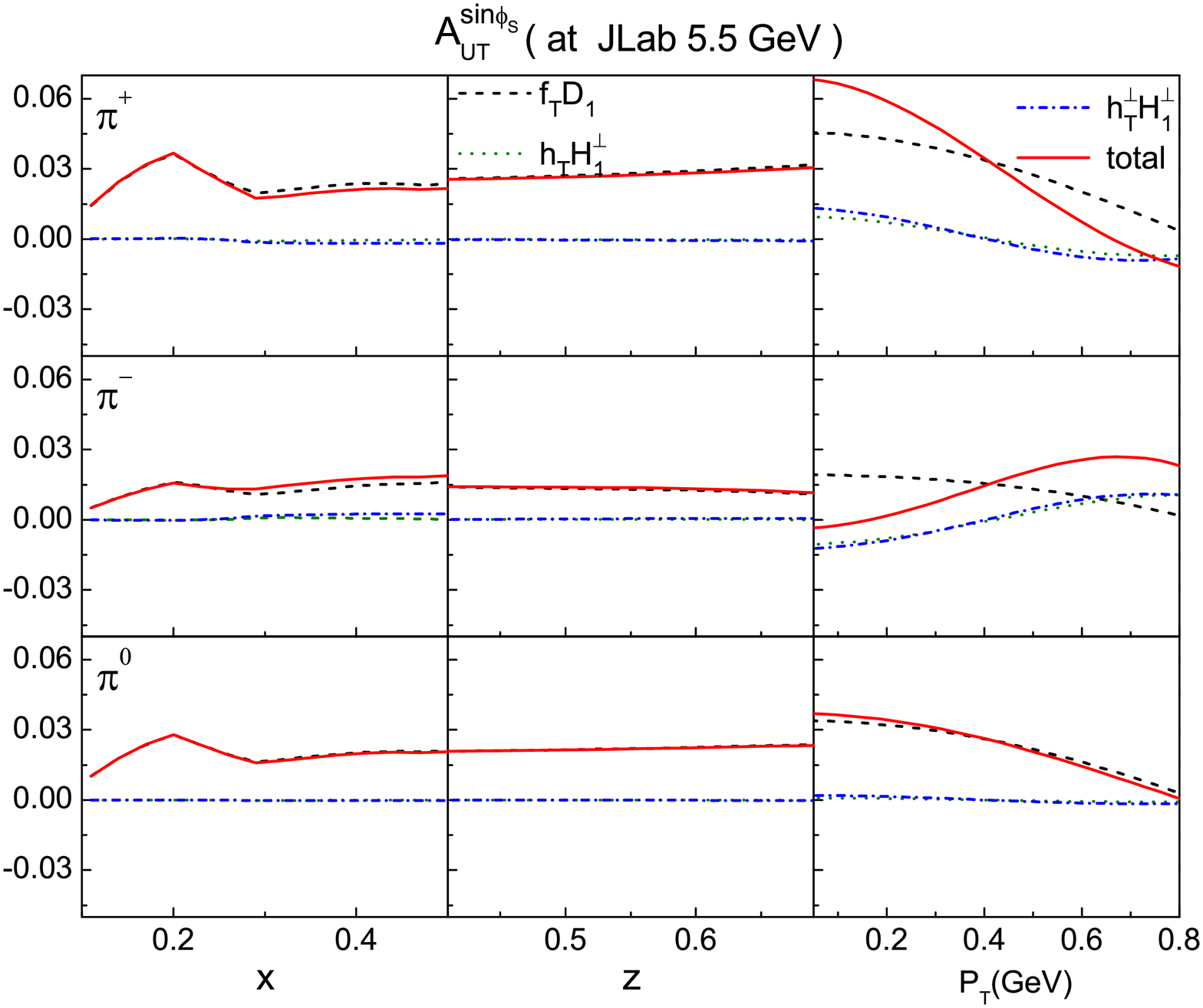}
   \includegraphics[width=0.99\columnwidth]{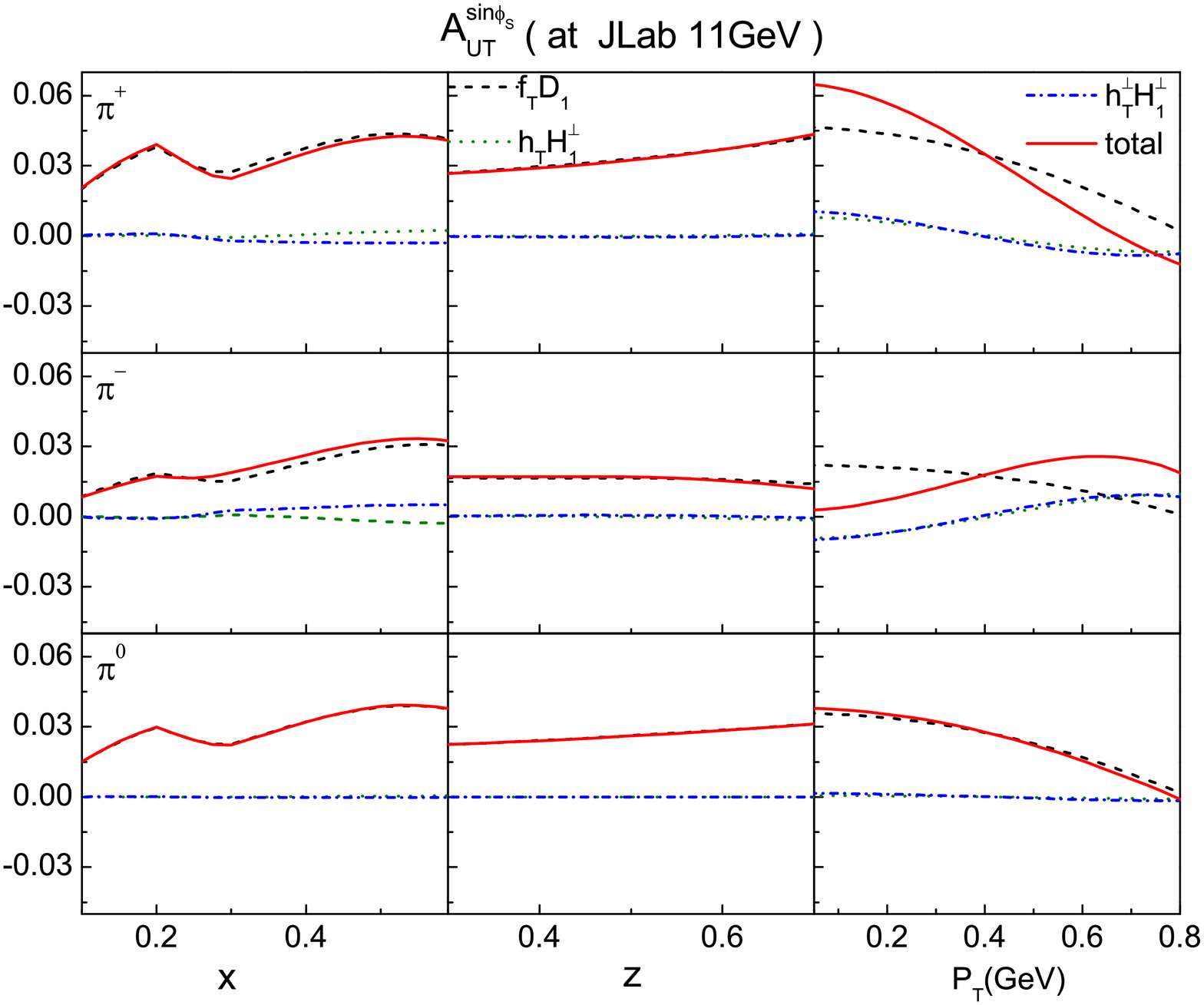}
  \caption{Prediction on the transverse SSA $A_{\mathrm{UT}}^{\sin\phi_S}$ for $\pi^+$ (upper panel), $\pi^-$ (middle panel) and $\pi^0$ (lower panel) in SIDIS at JLab with the beam energies $5.5\, \textrm{GeV}$ (upper) and $11\, \textrm{GeV}$ (lower).
  The dashed, dotted and dash-dotted curves represent the asymmetries from the $f_T D_1$, $h_T H_1^\perp$ and $h_T^\perp H_1^\perp$ terms, respectively. The solid curves correspond to the total contribution.}
  \label{JLaba}
  \end{figure}
 \begin{figure}
  \includegraphics[width=0.99\columnwidth]{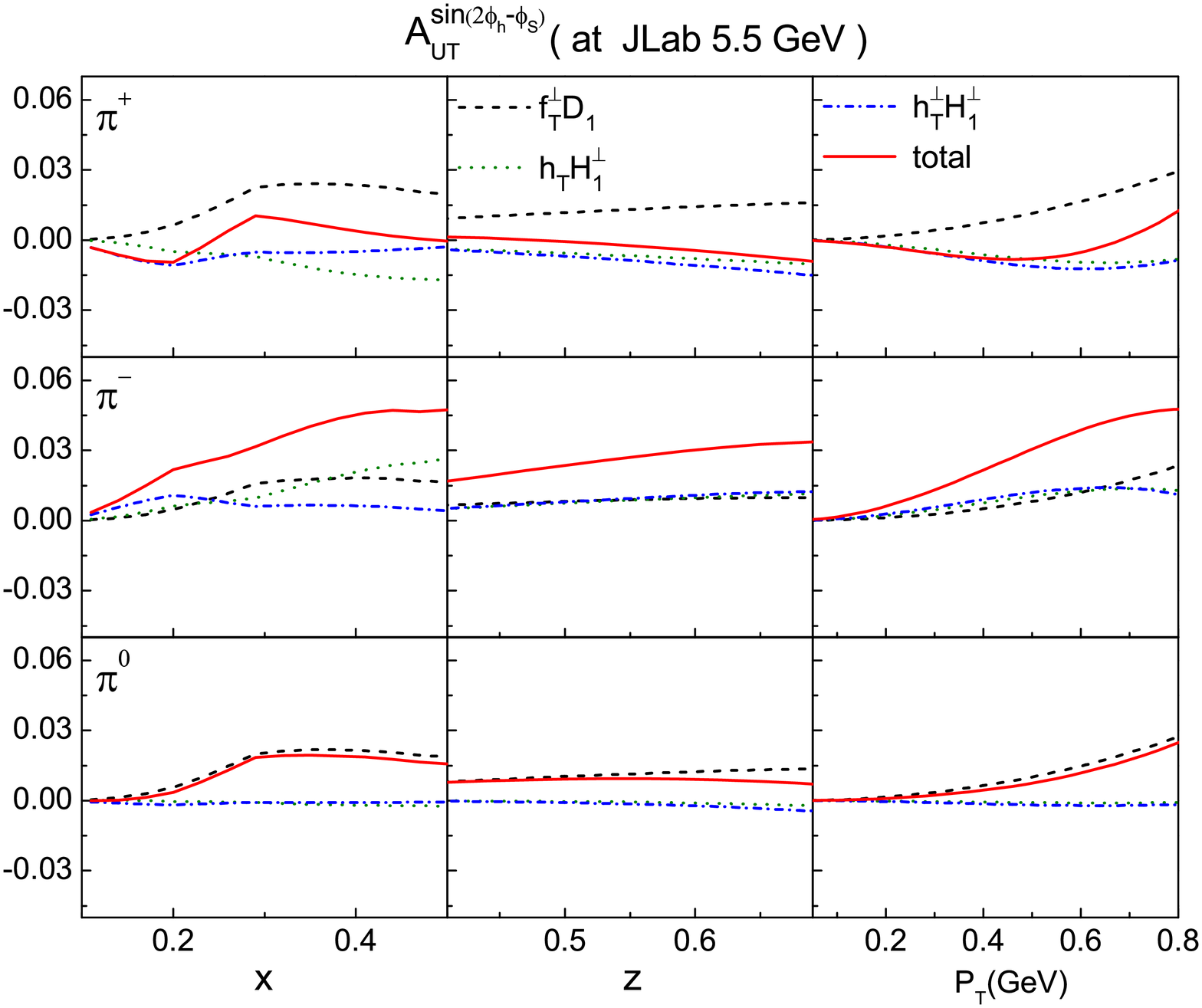}
   \includegraphics[width=0.99\columnwidth]{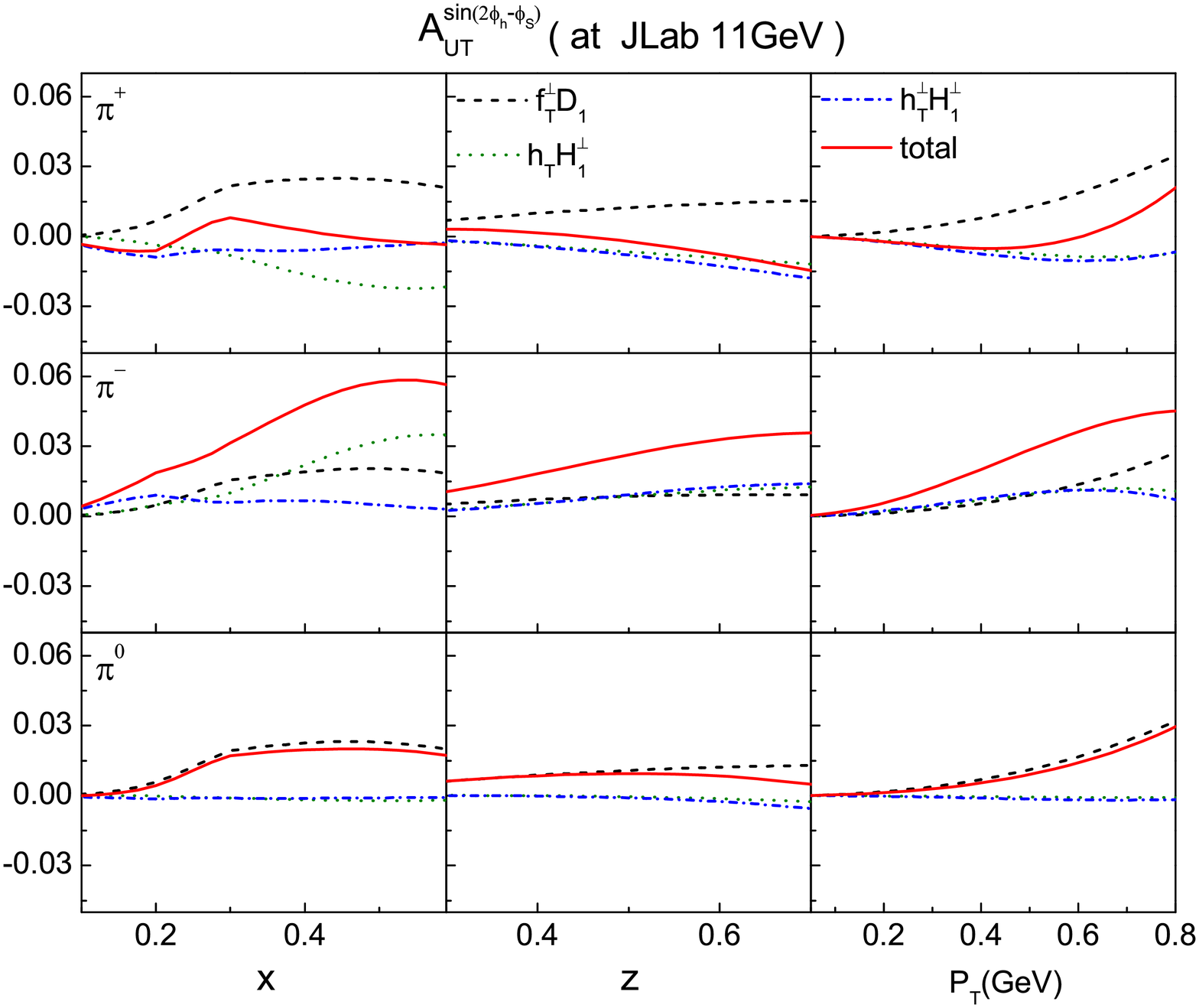}
  \caption{Similar to Fig.~\ref{JLaba}, but on the asymmetry $A_{\mathrm{UT}}^{\sin(2\phi_h -\phi_S)}$.
  The dashed, dotted and dash-dotted curves represent the asymmetries from the $f_T^\perp D_1$, $h_T H_1^\perp$ and $h_T^\perp H_1^\perp$ terms, respectively. The solid curves correspond to the total contribution.}
  \label{JLabb}
  \end{figure}

In Fig.~\ref{HERMESb}, we plot the prediction on the SSA $A_{\mathrm{UT}}^{\sin(2\phi_h -\phi_S)}$ vs $x$, $z$, and $P_T$.
In this case, generally, it is the $f_T^\perp D_1$ term that gives the main contribution, especially for the $x$- and $P_T$-dependent asymmetries.
The results show that nonzero asymmetry may be observed at $0.2<x<0.4$ or large $P_T$ region, where the size of the asymmetry is the largest.
Similar to the case of the SSA $A_{\mathrm{UT}}^{\sin \phi_S}$, we also find that the contributions from the  $h_T H_1^\perp$ and $h_T^\perp H_1^\perp$ to the $\pi^0$ asymmetry are consistent with zero.
According to Eq.~(\ref{collins}), i.e.,  Collins function for $\pi^0$, this result can be understood from the fact that the favored and unfavored Collins functions are similar in size but opposite in sign.

Here some comments are in order. In the $P_T$-dependent asymmetries in Figs.~\ref{HERMESa} and \ref{HERMESb}, we have plotted the curves up to $P_T\sim 1\,\rm{GeV}$, which is close to the typical hard scale $Q$ in the low energy SIDIS experiments.
Since the TMD-type formalism is only valid in the region $P_T\ll Q$, we admit that using the TMD formalism to predict the asymmetries at $P_T\sim 1\,\rm{GeV}$  may not be a good choice.
Nevertheless, we still show the results at large $P_T$ in Figs.~\ref{HERMESa} and \ref{HERMESb} for the possible comparison with data in the future.

\subsection{JLab $5.5 \,\textrm{GeV}$ and $11\,\textrm{GeV}$}

To test the feasibility to measure the transverse SSAs $A_{\mathrm{UT}}^{\sin \phi_S}$ and $A_{\mathrm{UT}}^{\sin(2\phi_h -\phi_S)}$ at the kinematics available at JLab, we also estimate them in the SIDIS with a $5.5 \,\textrm{GeV}$ electron beam, scattered off the transverse polarized proton, and we adopt the following kinematics in the calculation:
\begin{align}
&0.1<x<0.6,~~ 0.4<z<0.7,~~ Q^2>1\, \textrm{GeV}^2,\nonumber\\
&P_T>0.05\,\textrm{GeV},~~ W^2>4\,\textrm{GeV}^2.
\end{align}

In the upper panel of Fig~\ref{JLaba}, we plot our estimate on $A_{\mathrm{UT}}^{\sin\phi_S}$ at beam energy $5.5 \textrm{GeV}$ at JLab for $\pi^+$, $\pi^-$ and $\pi^0$ as functions of $x$, $z$, and $P_T$, respectively.
We find that the magnitude of the asymmetry $A_{\mathrm{UT}}^{\sin\phi_S}$ for $\pi^+$ and $\pi^0$ can reach $3\%$, which is sizable,
and again the $f_T D_1$ term dominates over the $h_T H_1^\perp$ and $h_T^\perp H_1^\perp$ terms.
The asymmetries as functions of  $x$ and $z$ are positive, while the asymmetry vs $P_T$ is positive at lower $P_T$ and turns to be negative at higher $P_T$, which is similar to tendency at HERMES.
This coincides with the $k_T$-dependence of the distribution $f_T$.

Similarly, we plot the results for $A_{\mathrm{UT}}^{\sin(2\phi_h -\phi_S)}$ in the upper panel of Fig~\ref{JLabb}.
Contrary to the asymmetry $A_{\mathrm{UT}}^{\sin\phi_S}$, in the case of $A_{\mathrm{UT}}^{\sin(2\phi_h -\phi_S)}$, the estimate shows that sizable asymmetry for $\pi^-$ may be observed at JLab.
Except the $\pi^0$ production, the contributions of the $h_T H_1^\perp$ and $h_T^\perp H_1^\perp$ terms to $A_{\mathrm{UT}}^{\sin(2\phi_h -\phi_S)}$ cannot be ignored.
The asymmetry for $\pi^+$ is small due to the cancelation between the positive $f_T D_1$ term and the negative $h_T H_1^\perp$ and $h_T^\perp H_1^\perp$ term, although individually their sizes are not small.

As an $11 \,\textrm{GeV}$ electron beam will be available at JLab very soon after the energy upgrading, for completeness we also predict the $\sin\phi_S$ and $\sin(2\phi_h-\phi_S)$ asymmetries at the following kinematics
\begin{align}
&0.08<x<0.6,~~ 0.2<y<0.9,\nonumber\\
&0.3<z<0.8,~~ Q^2>1\, \textrm{GeV}^2,\nonumber\\
& W^2>4\,\textrm{GeV}^2,~~ 0.05 <P_T<0.8\,\textrm{GeV}.
\end{align}
The results are shown in the lower panels of Figs.~\ref{JLaba} and ~\ref{JLabb}.
We find that the sizes and signs of the asymmetries at $11 \textrm{GeV}$ are similar to the results of $5.5$ GeV.

\subsection{COMPASS}

\begin{figure}
  \includegraphics[width=0.99\columnwidth]{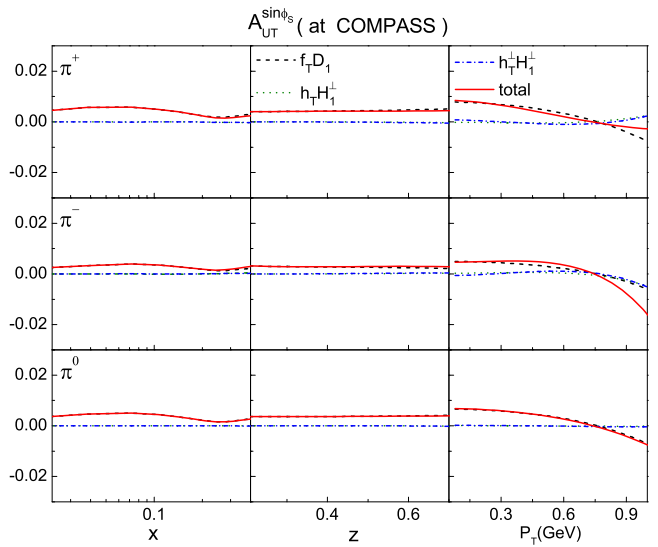}
  \caption{Predictions on the transverse SSA $A_{\mathrm{UT}}^{\sin{\phi_S}}$ for $\pi^+$ (upper panel), $\pi^-$ (middle panel) and $\pi^0$ (lower panel) in SIDIS at COMPASS.
  The dashed, dotted and dash-dotted curves represent the asymmetries from the $f_T D_1$, $h_T H_1^\perp$ and $h_T^\perp H_1^\perp$ terms, respectively. The solid curves correspond to the total contribution.}
  \label{compassa}
\end{figure}

\begin{figure}
  \includegraphics[width=0.99\columnwidth]{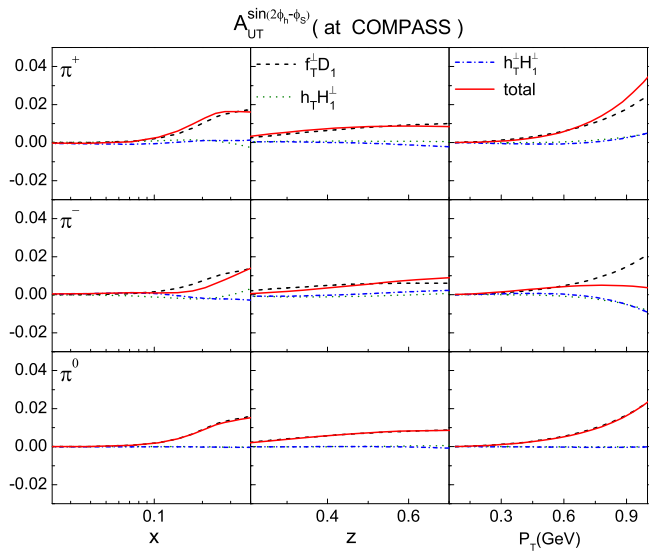}
  \caption{Similar to Fig.~\ref{compassa}, but on the asymmetry $A_{\mathrm{UT}}^{\sin{(2\phi_h -\phi_S)}}$.
  The dashed, dotted and dash-dotted curves represent the asymmetries from the $f_T^\perp D_1$, $h_T H_1^\perp$ and $h_T^\perp H_1^\perp$ terms, respectively. The solid curves correspond to the total contribution.}
  \label{compassb}
\end{figure}

For a further comparison, we also make the prediction on the transverse asymmetries at COMPASS with a muon beam of 160 GeV scattered off the proton target.
We show the results for the asymmetries $A_{\mathrm{UT}}^{\sin\phi_S}$ and $A_{\mathrm{UT}}^{\sin(2\phi_h -\phi_S)}$ in Figs.~\ref{compassa} and~\ref{compassb}, respectively.
In this calculation, we adopt the following kinematical cuts~\cite{Alekseev:2010rw}:
\begin{align}
&0.004<x<0.7,~~ 0.1<y<0.9,~~ z > 0.2,\nonumber\\
&P_T>0.1\,\textrm{GeV},~~Q^2>1\, \textrm{GeV}^2,\nonumber\\
&W>5\,\textrm{GeV}, ~~ E_h > 1.5\, \textrm{GeV}.
\end{align}

Our prediction shows that again the $f_T D_1$ term dominates the asymmetry $A_{\mathrm{UT}}^{\sin{\phi_S}}$, and its size is about $1\%$, which is clearly smaller than that at HERMES and JLab.
This is because the $Q^2$ at COMPASS ($1.3 \,\textrm{GeV}^2\,<Q^2\,<20.2\, \textrm{GeV}^2$~\cite{Alekseev:2010rw}) is larger than those at HERMES ($1.3 \,\textrm{GeV}^2<\,Q^2\,<6.2 \,\textrm{GeV}^2$~\cite{Airapetian:2004tw}) and JLab ($1.4\,\textrm{GeV}^2<\,Q^2\,<2.7 \,\textrm{GeV}^2$~\cite{Qian:2011py}), 
and the effect under study appears at subleading twist, thereby its size is suppressed by a factor $1/Q$.
In the case of the asymmetry $A_{\mathrm{UT}}^{\sin{(2\phi_h -\phi_S)}}$, the main contribution is from the $f_T^\perp D_1$ term (except the asymmetries for $\pi^+$ and $\pi^-$ in large $P_T$ region), similar to the case at HERMES.

\section{Conclusion}
\label{conclusion}

In this work, we studied the role of the twist-3 TMD distributions in the transverse SSAs.
We calculated the T-even twist-3 TMD distributions $h_T$ and $h_T^\perp$, together with the T-odd twist-3 TMD distributions $f_T$ and $f_T^\perp$, for the $u$ and $d$ valence quarks, in a spectator model with both the scalar and axial-vector diquarks.
In the calculation, we considered the differences between the isoscalar ($ud$-like) and the isovector ($uu$-like) spectators for the axial-vector diquark.
We employed the one-gluon exchange between the struck quark and the spectator to generate the T-odd structure, and chose the dipolar form factor for the nucleon-quark-diquark coupling to obtain finite results.
We also presented the flavor dependence of the four twist-3 TMD distributions as functions of $x$ and $k_T$, respectively.


Using the model results on the TMD distributions, for the first time we analyzed the transverse SSAs $A_{\mathrm{UT}}^{\sin\phi_S}$ and $A_{\mathrm{UT}}^{\sin(2\phi_h -\phi_S)}$ for charged and neutral pion production in SIDIS at the kinematics of HERMES, JLab and COMPASS.
We find that the estimated asymmetries are sizable, especially at HERMES and JLab, where the magnitude of the asymmetries can reach to 3 per cent at most.
Furthermore, the comparison between different origins of the asymmetries shows that the T-odd twist-3 TMD distributions play an important role in these asymmetries.
For the $\sin\phi_S$ asymmetry, the $f_T D_1$ term dominates in $\pi^+$, $\pi^-$ and $\pi^0$ production, while the $h_T H_1^\perp$ and $h_T^\perp H_1^\perp$ terms are almost negligible.
Also, the asymmetry from the $f_T D_1$ term tends to be positive at small $P_T$ region, while it turns to be negative in large $P_T$ region, due to the $k_T$-shape of the distribution $f_T$.
For the $\sin(2\phi_h -\phi_S)$ asymmetry, in the most cases, the main contribution is from the $f_T^\perp D_1$ term; the effects of the $h_T H_1^\perp$ and $h_T^\perp H_1^\perp$ terms might be observed in the asymmetry for $\pi^-$ production at JLab, according to our numerical calculation.

Based on the above results, we conclude that sizable $\sin\phi_S$ and $\sin(2\phi_h -\phi_S)$ asymmetries may be accessible at the kinematics of HERMES, JLab and COMPASS, by performing the SIDIS experiments on the transverse polarized proton target or analyzing the available data.
The measurements on the $P_T$-dependence of the asymmetry $A_{\mathrm{UT}}^{\sin\phi_S}$ may be employed to test the transverse momentum dependence of the distribution $f_T$, e.g., the existence of a node of $f_T$ in $k_T$.
Moreover, measuring the $\sin\phi_S$ and $\sin(2\phi_h -\phi_S)$ asymmetries for $\pi^0$ production, in which the contributions from $h_T$ and $h_T^\perp$ are negligible, are viable to provide clean probes on both the distributions $f_T$ and $f_T^\perp$.
Future experiments on these aspects can deepen our understanding on the role of twist-3 TMD distributions in transverse spin asymmetries.

\section*{Acknowledgements}
We are grateful to J.-P. Chen for bringing our attention on the subject of this paper.
This work is partially supported by National Natural Science
Foundation of China (Grant Nos.~11120101004 and 11035003), by the Fundamental Research Funds for the Central Universities (Grant No.~2242012R3007), and by the Qing Lan Project.
W. Mao is supported by the Scientific Research Foundation of the Graduate School of SEU (Grant No.~YBJJ1336) and by the Research and Innovation Project for College Postgraduate of Jiangsu Province (Grant No.~CXZZ13$\_$0079).

\end{document}